\documentclass[12pt,oneside,reqno,twoside]{article}

\usepackage{geometry,verbatim}
\usepackage[pdftex]{graphicx}

\usepackage[utf8]{inputenc}
\usepackage[OT2,T1]{fontenc}
\usepackage{comment} 

\usepackage{fancyhdr}

\usepackage{subcaption}

\usepackage[russian,ngerman,english]{babel}

\usepackage{
amsmath,
amssymb,
amsthm,
dsfont, 
mathabx, 
mathrsfs,
mathtools,
pgfplots
}
\usepackage{appendix}
\usepackage{wrapfig}
\usepackage{setspace}

\pgfplotsset{compat=1.18}

\usepackage{color}
\usepackage{braket}
\usepackage{oldgerm}
\usepackage{todonotes}

\usepackage{tikz}
\usetikzlibrary{trees}
\usetikzlibrary{decorations.pathmorphing}
\usetikzlibrary{decorations.markings}

\usepackage[colorlinks=true, linkcolor=blue, urlcolor=blue]{hyperref}

\mathtoolsset{showonlyrefs} 

\allowdisplaybreaks[1] 

\fancypagestyle{headings}{%
	\cfoot{\thepage}
}
\newfont{\suet}{suet14}
\newfont{\schwell}{schwell}
\DeclareTextFontCommand{\textsuet}{\suet}
\DeclareTextFontCommand{\textschwell}{\schwell}

\makeatletter
\newcommand{\proofstep}[2]{%
  \par
  \addvspace{\medskipamount}%
  \noindent\emph{Step #1: #2}\par\nobreak
  \addvspace{\smallskipamount}%
  \@afterheading
}
\makeatother

\numberwithin{equation}{section} 

\usepackage[shortlabels]{enumitem}

\renewcommand{\C}{\mathbb{C}} 
\newcommand{\R}{\mathbb{R}} 
\newcommand{\Z}{\mathbb{Z}} 
\newcommand{\N}{\mathbb{N}} 

\def\bx{\mathbf{x}}

\def\cH{\mathcal{H}}
\def\cF{\mathcal{F}}
\def\cE{\mathcal{E}}
\def\cO{\mathcal{O}}
\def\cS{\mathcal{S}}
\def\cQ{\mathcal{Q}}
\def\cW{\mathcal{W}}

\newcommand{\dx}[1]{ \, \mathrm{d}#1}
\newcommand{\dxx}[1]{\mathrm{d}#1}
\renewcommand{\d}{\mathrm{d}}

\def\vep{\varepsilon}

\newcommand{\e}[1]{\mathrm{e}^{#1}}
\renewcommand{\i}{\mathrm{i}}
\DeclareMathOperator{\Span}{span}

\DeclareMathOperator{\supp}{supp}
\def\sym{\operatorname{Sym}}  

\def\dist{\operatorname{dist}}

\DeclareMathOperator{\sgn}{sgn}

\newcommand{\scp}[2]{\left\langle #1,#2\right\rangle} 

\def\nb{\mathcal{N}}
\def\ad{a^\dagger}
\def\ca{\check{a}}
\def\cad{\check{a}^\dagger}
\def\ball{B}
\def\enum{n}

\def\EPT{E_{PT}}
\def\HPT{\cE_{PT} }
\def\Hph{\nb}

\swapnumbers
\theoremstyle{plain}
\newtheorem{prop}{Proposition}[section]

\newtheorem{lem}[prop]{Lemma}
\newtheorem{thm}[prop]{Theorem}
\newtheorem{cor}[prop]{Corollary}
\newtheorem*{cor*}{Corollary}
\newtheorem{rem}[prop]{Remark}
\newtheorem*{rem*}{Remark}

\newtheorem*{example*}{Example}

\theoremstyle{definition}

\newtheorem*{assum*}{Assumptions}

\usepackage{orcidlink}
\usepackage[affil-it]{authblk}

\begin{document}

\title{The Lieb--Thomas strategy for strongly coupled fermionic multipolarons with general external fields}

\author[1]{Ioannis Anapolitanos\thanks{E-mail: \texttt{ioannis.anapolitanos@kit.edu}}}
\affil[1]{Department of Mathematics, Karlsruhe Institute of Technology, Karlsruhe, Germany}
\author[2,3]{Michael Hott\orcidlink{0000-0003-4243-6585}%
		\thanks{E-mail: \texttt{mhott@odu.edu}}}
        
        \affil[2]{Simons-Laufer Mathematical Sciences Institute, Berkeley, CA 94720-5070, USA}
        \affil[3]{Department of Mathematics \& Statistics, Old Dominion University, Norfolk, VA 23529}

\newgeometry{margin=1in}

\maketitle

\begin{abstract}
    In this article, we prove that the ground-state energy of a fermionic Fr\"ohlich multipolaron can be approximated, in the strong electron-phonon coupling limit, by the ground-state energy of a corresponding fermionic Pekar-Tomasevich multipolaron, even in the presence of external electric and magnetic fields. Our analysis builds upon Lieb and Thomas' approach \cite{liebthomas}, which was originally developed for a single polaron without external fields, and Wellig's generalization to multipolarons \cite{wellig} with (specialized) external fields. Our main new contributions are twofold. First, we take into account the fermionic statistics of the multipolaron by employing a localization method from \cite{liebloss}. Second, we relax an assumption in \cite{wellig} on the external electric and magnetic fields, which is not easily verifiable unless the fields are periodic. Instead, we allow for general fields that only ensure self-adjointness of the Fr\"ohlich Hamiltonian. In particular, our work demonstrates the robustness of the Lieb--Thomas strategy when extended to fermionic multipolarons and general external potentials.
\end{abstract}

    \paragraph{Statements and Declarations} The authors do not declare financial or non-financial interests that are directly or indirectly related to the work submitted for publication.
	
	\paragraph{Data availability} The manuscript has no associated data.
	
\paragraph{Acknowledgments} The first author (IA) is grateful to Marcel Griesemer for suggesting that we work on the strong-coupling limit of polarons, to Jeremy Faupin for discussions on non-relativistic quantum electrodynamics that turned out to be inspiring for this work, and to Marcel Griesemer, Robert Seiringer, David Wellig, Benjamin Landon and Andreas W\"unsch for numerous discussions on polarons and the strong-coupling limit. Both authors gratefully acknowledge Dirk Hundertmark and Semjon Vugalter for numerous discussions and references that helped complete the arguments of the proofs. The research of (IA) was funded by the Deutsche Forschungsgemeinschaft (DFG, German Research Foundation) Project-ID 258734477 -- SFB 1173. This material is based on work supported by the National Science Foundation under Grant No. DMS-2424139; (MH) was in residence at the Simons Laufer Mathematical Sciences Institute in Berkeley, California, during the Fall 2025 semester.

\tableofcontents

\restoregeometry

\section{Introduction}

\subsection{Background}

\begin{figure}[b]
\centering
\hspace{\fill}
\begin{subfigure}{0.3\linewidth}
\centering
    \includegraphics[width=0.8\linewidth]{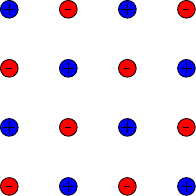}
\caption{Polar crystal}
\label{fig-polarcrystal}
\end{subfigure}
\hspace{\fill}
\begin{subfigure}{0.3\linewidth}
 \centering
\includegraphics[width=0.8\linewidth]{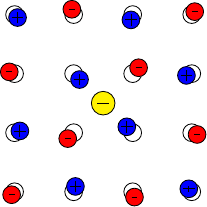} 
\caption{Polaron}
\label{fig-polaron}
\end{subfigure}
\hspace{\fill}
\caption{Ionic crystal with and without traveling test charge}
\end{figure}
 
Consider an electron moving through a polar crystal (Fig. \ref{fig-polarcrystal}). Such a crystal consists of alternating positive and negative ions; a standard example is NaCl (table salt). We assume the ions are not rigidly fixed. As the electron moves, it distorts the lattice in its immediate neighborhood (Fig. \ref{fig-polaron}). The electron together with this induced distortion is called a \emph{polaron}. For two or more electrons, the corresponding objects are \emph{bipolarons} or \emph{multipolarons}. We model the (quantized) lattice vibrations by a massless bosonic field; its excitations are \emph{phonons}, and electrons interact with them through the field. 
\par It is energetically favorable for electrons to deform the lattice in a localized region, which can keep them close together. The resulting distortion produces an effective attraction that competes with Coulomb repulsion; thus polarons may attract even though bare electrons repel. To study effective multipolaron interactions, we focus on the ground-state energy. A widely accepted model is the \emph{Fr\"ohlich model}, proposed by Herbert Fr\"ohlich \cite{froehlich} in 1954; see also \cite{callaway}. In the same year, Pekar \cite{pekar} argued that in the strong-coupling limit the phonon modes can be integrated out; see also \cite{leelowpines}. Together with Tomasevich \cite{pekartomasevich}, this yields the effective \emph{Pekar-Tomasevich multipolaron} model, with the single-polaron case known as the \emph{Pekar polaron}. Feynman \cite{feynman1955slow,feynman} used a Feynman--Kac formula to expand the Fr\"ohlich ground-state energy in the coupling constant and identified the leading term as the Pekar polaron energy. 

\par In 1980, Adamowski, Gerlach and Leschke \cite{leschke} sketched a proof of convergence of the Fr\"ohlich polaron ground-state energy to the Pekar-Tomasevich energy using large deviations for Brownian motion. In 1983, Donsker and Varadhan \cite{donsker} gave a complete proof, but without rates. In 1997, Lieb and Thomas \cite{liebthomas} provided a simpler argument that also yields convergence rates. Their approach applies Pekar's idea to a UV-truncated and spatially localized Fr\"ohlich model and controls the truncations in the strong-coupling limit. We adapt their strategy, which proceeds in the following steps.

\begin{enumerate}[(i)]
    \item \textbf{Electron localization:} Up to an energy error $\Delta E$, electrons can be localized inside a cube of side-length $\sim(\Delta E)^{-\frac12}$, in accordance with Heisenberg's uncertainty principle.
    \item \textbf{UV cutoff:} Using a commutator estimate, see Lemma \ref{lem-UV-cutoff}, they truncate UV phonon modes with momentum above $\Lambda$. 
    
    \item \textbf{Block-mode reduction:} To reduce the number of phonon modes, they dissect the $\Lambda$-cube into $(\Lambda/P)^3$ blocks of side length $P$; see Prop. \ref{prop-block-partition}. Using translation invariance of the Hamiltonian, the interaction per block reduces to that of an electron with a single representative block mode in the cube.
    
    \item \textbf{Phonon integration:} Using coherent state analysis, they then employ Pekar's argument, which is based on completion of the square, to this reduced Hamiltonian, see Prop. \ref{prop-block-mode-integration} and, for a more detailed explanation, Appendix \ref{app-PT}.
\end{enumerate} 
Both the many-particle case and the presence of external potentials introduce new challenges. With many particles, interparticle distances create an additional length scale that can interfere with the localization step. External potentials break translation invariance, which underlies the block reduction. Moreover, the final Lieb--Thomas reduction uses the scale invariance of the Pekar functional, which is violated by both the Coulomb interaction and external fields. 

\par In 2013, Wellig and Griesemer \cite{griesemer-wellig-strong-polaron-static-fields} studied the single polaron with external electric and magnetic fields. In the same year, the first author and Landon \cite{anapolandon} generalized the Lieb--Thomas argument to the multipolaron case without external fields. To handle many particles, they replaced the electronic localization step with 
\begin{enumerate}[(i')]
    \item\label{itm:cluster-loc} \textbf{Cluster localization:} Split the multipolaron into various localized clusters.
\end{enumerate}
Inspired by \cite{frank-lieb-seiringer-thomas-stability-absence}, they use Feynman--Kac formulas to bound the energy of interacting clusters by that of non-interacting clusters. To close the Lieb--Thomas argument, they add:
\begin{enumerate}[(i)]
\setcounter{enumi}{4}
    \item\label{itm-eff-total-cluster} \textbf{Effective collective cluster energy:} Bound the sum of Pekar-Tomasevich energies of the clusters by the total energy of the Pekar-Tomasevich multipolaron, see Corollary \ref{cor-EPT-B-subadd}.
\end{enumerate}
We denote by $\EPT^{(n,\alpha)}$ the ground-state energy of the Pekar--Tomasevich $n$--polaron with coupling constant $\alpha$ between the electrons and the phonon field.
In the absence of external fields, the Pekar-Tomasevich functional is translationally invariant and consequently obeys the subadditivity
\begin{equation}\label{eq:asswellig}
	\EPT^{(n,\alpha)} \leq \EPT^{(k,\alpha)} + \EPT^{(n-k,\alpha)}, \quad \forall 1\leq k < n \leq N \, .
\end{equation}
We explain this in more detail in Rem. \ref{rem-EPT-subadd-TI}. As a corollary of strong-coupling convergence, they show that bipolarons form a bound state if the effective Coulomb coupling $\nu$ is below a threshold $\nu_c>2$; see \eqref{def-Hamiltonian}. 
\par A few months later, Wellig \cite{wellig} extended \cite{anapolandon} to include external potentials. A key difficulty is that the Feynman--Kac formula is no longer available, and \eqref{eq:asswellig} need not hold. In the field-free case, Feynman--Kac allows simultaneous localization of electronic support and interacting phonon modes. Instead, Wellig first proves a covering lemma (Lemma \ref{lem-covering-lemma}) and develops a cluster localization argument similar to \cite{anapolandon} (see Prop. \ref{prop-cluster-localization}). He then generalizes the proof to arbitrary clusters by localizing phonons to regions near the electron clusters they interact with, following \cite{frank-lieb-seiringer-thomas-stability-absence}; see Lemma \ref{prop-block-partition}. To carry out step \ref{itm-eff-total-cluster}, he assumes \eqref{eq:asswellig} for the external fields; periodic fields provide a notable example. 

\par Note that neither \cite{anapolandon} nor \cite{wellig} allow for fermionic statistics, since their localization functions do not preserve particle statistics. We overcome this challenge by implementing localization functions used in \cite{liebloss} in the cluster localization argument \ref{itm:cluster-loc}, at the cost of larger error terms. Our error is of order $O(N^{\frac{82}{30}})$ as $N\to\infty$, while the error presented in \cite{wellig} grows as $O(N^3)$; however, optimizing parameters there leads to an $O(N^{\frac{57}{23}})$ error. To remove \eqref{eq:asswellig}, we exploit the fact that cluster localization can be preserved throughout the phonon integration, i.e., in the reduced Pekar-Tomasevich model, in the Lieb--Thomas argument. Since the clusters are well separated, their interaction can be controlled, allowing us to replace \eqref{eq:asswellig} with the estimate
\begin{align}
    \EPT^{(N,\alpha)} \, \leq \, \sum_{j=1}^m\EPT^{(n_j,\alpha)}(B_j)+C\frac{N^2\alpha}R, \quad \sum_{j=1}^mn_j=N \, , \quad \min_{j\neq k}\dist(B_j,B_k) \geq R \, ,
\end{align}
see Corollary \ref{cor-EPT-B-subadd}. Here $\EPT(B_j)$ denotes the Pekar-Tomasevich ground-state for multipolarons confined to a ball $B_j\subseteq \R^3$. This estimate holds for very general external potentials and imposes no additional constraints beyond self-adjointness of the Fr\"ohlich Hamiltonian.

\paragraph{Further directions} Miyao and Spohn \cite{miyaospohn} proved strong resolvent convergence of the UV cut-off Fr\"ohlich Hamiltonian to the full model and studied existence of the ground-state energy at fixed total momentum. Ghanta \cite{Ghanta} showed that in one dimension with a single electron ($N=1$), no magnetic field ($A=0$), and suitable assumptions on $V$, approximate minimizers of the Fröhlich Hamiltonian converge (up to rotation and translation) to the unique minimizer of the Pekar-Tomasevich functional in an appropriate topology. We do not address this convergence here.
\par Another active direction concerns the effective mass of the Fr\"ohlich polaron, which increases relative to the uncoupled electron mass \cite{bazaes2025effectivemassfrohlichpolaron,brooks2024prooflandaupekarformulaeffective,sellke-almost-qaurtic-lb-effective-mass-froehlich-polaron}, and the related dependence of the energy on total momentum \cite{brooks-seiringer-froehlich-polaron-strong-coupling-II,lampart-mitrouskas-global-min-energy-momentum-relation-polaron,mitrouskas-krzysztof-optimal-UB-energy-momentum,polzer-renewal-approach-froehlich-polaron}. Brooks and Seiringer \cite{brooks-seiringer-froehlich-polaron-strong-coupling-I} studied the associated Bogoliubov theory. For asymptotic low-energy expansions, see \cite{brooks-mitrouskas-low-energy-asymptotics,seiringer-polaron-strong-coupling}. We do not pursue these directions here, but extending our ideas to them would be interesting.

\subsection{Definition of models}

\begin{figure}
    \centering
    \hspace{\fill}
    \begin{subfigure}[t]{0.46\linewidth}
    \centering
    \includegraphics[width=\linewidth]{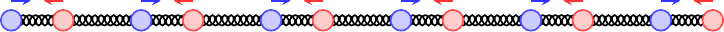}
    \caption{Optical Phonons in a diatomic 1D chain. Atoms of species $A$ and $B$ move out of phase}
    \label{fig-optical-phonons}
    \end{subfigure}
    \hspace{\fill}
    \begin{subfigure}[t]{0.46\linewidth}
    \centering
    \includegraphics[width=\linewidth]{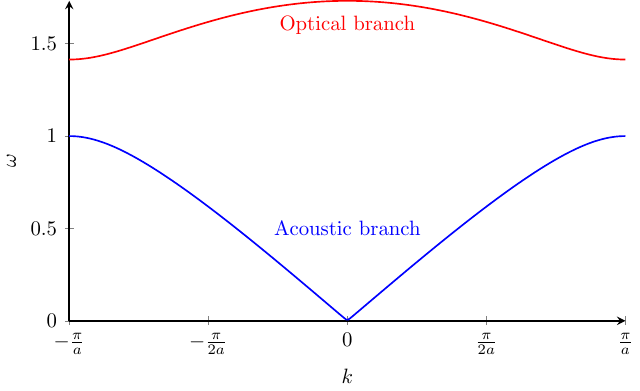}
    \caption{Phonon dispersion in diatomic 1D chain with lattice constant $a>0$}
    \label{fig-phonon-dispersion}
    \end{subfigure}
    \hspace{\fill}
    \caption{Definition of phonon modes}
\end{figure}

Lattice vibrations with neighboring atoms moving out of phase (e.g., in a diatomic crystal) are called \emph{optical phonons}; see Fig. \ref{fig-optical-phonons}. In three dimensions, phonon momentum can be aligned with or orthogonal to the electron momentum. We focus on the aligned case, i.e., \emph{longitudinal optical phonons}. Near the origin, the dispersion is nearly constant; see Fig. \ref{fig-phonon-dispersion} for the one-dimensional case. This constant is the \emph{Debye frequency}, which we set to 1. We work in the continuum limit (lattice spacing $\to 0$), though the analysis can be repeated in a discrete setting as in \cite{liebthomas}. The resulting model of electrons coupled to longitudinal optical phonons in a polar crystal is the \emph{Fr\"ohlich model} \cite{froehlich}.

\par Given $N\in\N$, $\nu\in\R$, $\alpha>0$, the Fr\"ohlich Hamiltonian is defined by
\begin{align}\label{def-Hamiltonian}
	H^{(N,\alpha)} \, &= \, \sum_{j=1}^N\Big(D_{A_\alpha,x_j}^2+V_\alpha(x_j)+\sum_{k=j+1}^N\frac{\nu\alpha}{|x_j-x_k|}\Big)\\
    &\qquad +\frac{\sqrt{\alpha}}{\sqrt{2}\pi} \sum_{j=1}^N\int_{\R^3}\left(a(k)\e{\i kx_j}+\ad(k)\e{-\i kx_j}\right)\frac{\d k}{|k|}+\int_{\R^3}\ad(k)a(k)\dx{k} \, .
\end{align}
Here, we let $A\in L^2_{\mathrm{loc}}(\R^3;\R^3)$ and $V:\R^3\to\R$ be an external magnetic and electric potential, respectively. We assume that $V$
is relatively $-\Delta$-form bounded with relative form bound $0$, i.e., for all $\vep>0$ there exists $C_\vep>0$ s.t.
\begin{align}
    |\scp{\Psi}{V\Psi}| \, \leq \, \vep \|\nabla \Psi\|_2^2 \, + \, C_\vep \|\Psi\|_2^2 \quad \forall \Psi\in L^2(\R^3) \, .
\end{align}
The rescaled external potentials are defined by
\begin{align}\label{eq-rescaled-fields}
    A_\alpha(x) \, := \, \alpha A(\alpha x), \quad V_\alpha(x) \, := \, \alpha^2V(\alpha x) \, .
\end{align}
For any $\tilde{A}\in L^2_{\mathrm{loc}}(\R^3;\R^3)$, we denote the magnetic gradient
\begin{align}\label{magnetic-gradient}
    D_{\tilde{A},x} \, := \, -i\nabla_x+\tilde{A}(x) \, .
\end{align}
The assumption $A_k\in L^2_{\mathrm{loc}}(\R^3)$, $k=1,2,3$ ensures that $D_{A_\alpha,x}$ with domain
\begin{align}
    H^1_{A_\alpha}(\R^3) \, := \, \{f\in L^2(\R^3)\mid  \forall k=1,2,3:\, (-i\partial_{x_k} +(A_\alpha)_k)f\in L^2(\R^3)\} 
\end{align}
is self-adjoint, see \cite[Section 7.20-7.22]{llbook}.
\par Electrons are described by the fermionic Hilbert space $\bigwedge_{j=1}^N L^2(\R^3)$ of wavefunctions $\Psi$ satisfying for almost all $x_1,\ldots,x_N\in\R^3$ and all permutations $\pi\in\cS_N$ of labels $\{1,\ldots,N\}$ the equality
\begin{align}\label{def-fermionic-wavefunction}
    \Psi(x_{\pi(1)},\ldots,x_{\pi(N)}) \, = \, \sgn(\pi)\Psi(x_{1},\ldots,x_{N}) \, . 
\end{align}
$a$, $\ad$ are the bosonic creation and annihilation operators on the bosonic Fock space 
\begin{equation}\label{def-fock}
    \cF \, := \, \bigoplus_{n=0}^\infty\bigvee_{j=1}^n L^2(\R^3) \, ,    
\end{equation}
where $\bigvee_{j=1}^n L^2(\R^3)$ denotes the \emph{symmetric} subspace of $L^2(\R^{3n})$. An element $\eta_n\in \bigvee_{j=1}^n L^2(\R^3)$ satisfies almost everywhere the equality
\begin{align}\label{def-bosonic-wavefunction}
    \eta_n(x_1,\ldots,x_n) \, = \, \eta_n(x_{\pi(1)},\ldots,x_{\pi(n)}) \quad \forall \pi\in \cS_n \, .
\end{align}
$a(k)$, $\ad(\ell)$ satisfy the \emph{canonical commutation relations (CCR)}
\begin{equation}\label{CCR}
	[a(k),\ad(\ell)]=\delta(k-\ell) \, , \qquad [a(k),a(\ell)] \, = \, [\ad(k),\ad(\ell)] \, = \, 0 \, .
\end{equation}	
The electron-phonon coupling is given by
\begin{equation}
	\phi(x)\;:=\;\frac{1}{\sqrt{2}\pi}\int_{\R^3}\left(a(k)\e{\i kx}+\ad(k)\e{-\i kx}\right)\frac{\d k}{|k|} \, .
\end{equation}
It describes electron scattering by a phonon with momentum transfer $k\in\R^3$. More precisely, annihilation of a phonon of momentum $k$ results in an electron gaining momentum $k$. Creation of a phonon of momentum $k$ results in an electron losing momentum $k$. The form factor $(\sqrt{2}\pi|k|)^{-1}$ is tied to the scattering of an electron with momentum $k$ by longitudinally polarized optical phonons. It is a product of the electron momentum $|k|$, the Fourier transform of the Coulomb potential $(2\pi^2|k|^2)^{-1}$, and the phonon amplitude $\pi\sqrt{2 \omega(k)}$. The dispersion $\omega$ of longitudinal optical phonons is approximately constant -- its value is known as the \emph{Debye frequency} -- and units are set such that $\omega(k)=1$. For more details, we refer to \cite[Chapter 8.1]{feynman}.
\par Since the dispersion $\omega(k)=1$ is constant, the phonon field energy is given by the number of excitations, namely
\begin{equation}
	\nb \, := \, \int_{\R^3}\ad(k)a(k)\dx{k} \, .
\end{equation} 
The total Hilbert space describing both phonons and electrons is given by
\begin{equation}
	\cH_N \, := \, \Big(\bigwedge_{j=1}^N L^2(\R^3)\Big)\otimes\cF \subseteq \Big(\bigotimes_{j=1}^N L^2(\R^3)\Big)\otimes\cF \cong L^2(\R^{3N};\cF) \, . \label{def-hilbert}
\end{equation}
We will refer to $\Psi\in\cH_N$ as a \emph{multipolaron}, which describes the ensemble of $N$ electrons and a phonon field. 
\par A form core of $H^{(N,\alpha)}$ is given by
\begin{equation}\label{eq:qndef}
	 \cQ_0 \, := \, \Big(\bigwedge_{j=1}^N C_c^\infty(\R^3)\Big)\otimes\mathcal{F}_0 \, ,
\end{equation}
where 
\begin{equation}
\mathcal{F}_0 \, := \, \{(\eta_n)_{n\in\N}\in\mathcal{F}|\;\forall n\in\N: \eta_n\in \bigvee_{j=1}^n C_c^\infty(\R^3)
\;\exists n_0\in\N:\eta_n=0\;\forall n\geq n_0\} \, .
\label{def:f0}
\end{equation}
A difficulty in realizing $H^{(N,\alpha)}$ as a self-adjoint operator stems from the fact that $\frac{e^{ix\cdot}}{|\cdot|}\notin L^2(\R^3)$. To overcome this obstacle, one casts $H^{(N,\alpha)}$ as a quadratic form $q^{(N,\alpha)}$ on $\cQ_0$ and proves that it is closable. In fact, $\phi$ is infinitesimally relatively form-bounded w.r.t. $D_{\tilde{A}}^2+\nb$. Then the KLMN theorem, see \cite[Theorem X.17, p. 167]{reed-simon-II-fourier-selfadjoint}, \cite[Theorem 3.19]{lewin-spectraltheory+QM-book}, yields that $q^{(N,\alpha)}$ has a unique closed extension on $\cH_N$. This closed extension then is associated with a unique self-adjoint operator that is bounded from below, see \cite[Theorem 2, Chapter 10]{birsol}. For more details, we refer to \cite{hott,miyaospohn,nelson}. For a description of the domain of self-adjointness, we refer to \cite{griesemer-wuensch-froehlich-sa}.
\par By the Courant-Fischer min-max principle, the ground-state energy $E^{(N,\alpha)}:=\inf\sigma(H^{(N,\alpha)})$ satisfies  
\begin{equation}
E^{(N,\alpha)} \; = \; \inf_{\Phi_N\in\cQ_0 , \|\Phi_N\|=1}\scp{\Phi_N}{H^{(N,\alpha)}\Phi_N}.
\label{eq:energydef}
\end{equation}
The Pekar-Tomasevich functional $\HPT^{(N,\alpha)}$ is given for $\Psi_N\in \bigwedge_{j=1}^N C_c^\infty(\R^3)$ by
\begin{equation}
\HPT^{(N,\alpha)}(\Psi_N) \, = \, \inf_{\substack{\eta\in\mathcal{F}_0,\\\|\eta\|=1}} \scp{\Psi_N\otimes\eta}{H^{(N,\alpha)}\Psi_N\otimes\eta}.\label{eq:ptdef}
\end{equation}
Let 
\begin{align}\label{def-density}
    \rho_{\Psi_N}(x) \, := \, \sum_{j=1}^N \int_{\R^{3(j-1)}}\int_{\R^{3(N-j)}}|{\Psi_N}(\bx_{j-1},x,\bx_{N-j}')|^2 \dx{\bx_{N-j}'}\dx{\bx_{j-1}}  \, .
\end{align}
denote the density of $\Psi_N\in \bigwedge_{j=1}^NL^2(\R^3)$. A now standard calculation originally found in \cite{pekar}, see Lemma \ref{lem-pekar}, implies that
 \begin{align}
    \HPT^{(N,\alpha)}(\Psi_N) \, &= \, \sum_{j=1}^N\scp{\Psi_N}{\Big(D_{A_\alpha,x_j}^2+V_\alpha(x_j)+\sum_{k=j+1}^N\frac{\nu\alpha}{|x_j-x_k|}\Big)\Psi_N}\\
    &\qquad -\alpha \int \frac{\rho_{\Psi_N}(x)\rho_{\Psi_N}(y)}{|x-y|}\dx{x}\dx{y}.
 \end{align}
We then denote the ground-state energy in the Pekar-Tomasevich model by 
\begin{equation}\label{def-EPT}
    \EPT^{(N,\alpha)} \, := \, \inf_{\substack{\Psi_N\in \bigwedge_{j=1}^N C_c^\infty(\R^3) ,\\\|\Psi_N\|=1}}\HPT^{(N,\alpha)}(\Psi_N).
\end{equation}
As an immediate consequence, we have that
\begin{equation}
E^{(N,\alpha)} \, \leq \, \EPT^{(N,\alpha)}. \label{ineq:ec}
\end{equation}

\begin{rem}[Homogeneity of $\EPT^{(N,\cdot)}$]\label{rem-PTscale}
	By rescaling $\Psi_\alpha(x)=\alpha^{\frac{3N}{2}}\Psi(\alpha x)$, we obtain that
	\begin{equation}\label{eq:resc}
	\HPT^{(N,\alpha)}(\Psi_\alpha) \, = \,\alpha^2\HPT^{(N,1)}(\Psi) \, .
	\end{equation}
	Consequently, the Pekar-Tomasevich energy satisfies $\EPT^{(N,\alpha)}=\alpha^2\EPT^{(N,1)}$. 
\end{rem}

\subsection{Main result}
 
 Our goal is to show that the Pekar-Tomasevich functional, which is a simpler model than the full Fröhlich Hamiltonian, yields the behavior of the ground-state energy of the Fröhlich Hamiltonian up to a relative error which vanishes in the strong-coupling limit $\alpha\to\infty$. In view of \eqref{ineq:ec}, an appropriate lower bound for the ground-state energy of the Fröhlich Hamiltonian is sufficient. Such a lower bound is provided in our main result, Theorem \ref{thm:mainthm}. 

\begin{thm}[Validity of Pekar-Tomasevich reduction in strong-coupling limit]\label{thm:mainthm}
	Let $V$ be relatively $-\Delta$-form-bounded with relative form bound $0$, $A_k\in L^2_{\mathrm{loc}}(\R^3)$, $k=1,2,3$, and $\nu\in\R$. Then there exists a constant $C\in\R$ such that for any $N\in\N$, and $\alpha\geq1$ we have that
		\begin{equation}\label{eq-main-thm}
		E^{(N,\alpha)} \, \geq \, \EPT^{(N,\alpha)}-CN^\frac{82}{30}\alpha^{\frac{42}{23}} \, .
		\end{equation}
        In particular, we have that
        \begin{align}
            \lim_{\alpha\to\infty}\frac{E^{(N,\alpha)}}{\alpha^2} \, = \, \EPT^{(N,1)} \, .
        \end{align}
\end{thm}

\begin{rem}[Dependence of constant $C$]\label{rem-dependence-on-constant}
    The constant $C$ in Theorem \ref{thm:mainthm} only depends on $|\nu|$ and
    \begin{align}
    \MoveEqLeft \inf_{\substack{N\in\N,\\ \lambda\in\{1,2\}}}\frac1{N^2}\inf_{\substack{\Psi_N\in \C^\infty_c(\R^3)\\\|\Psi_N\|=1}}\bigg(\langle\Psi_N,\Big(\sum_{j=1}^N\big(D_{A,x_j}^2+\lambda V(x_j)\big)+ \sum_{k=j+1}^N\frac{\lambda\nu}{|x_j-x_k|}\Big)\Psi_N\rangle \\
    &\qquad -\lambda^2 \int\frac{\rho_{\Psi_N}(x)\rho_{\Psi_N}(y)}{|x-y|} \dx{x}\dx{y} \bigg), \label{eq-EPT-sup-bound}
    \end{align}
    see Lemma \ref{lem-concavity}. The infimum in \eqref{eq-EPT-sup-bound} is finite, since the electron-electron interaction grows at most with $N^2$, while the rest grows to linear order. To bound the terms, one employs Hardy's inequality together with the diamagnetic inequality \cite[Theorem 7.21]{llbook}, and the assumptions on $A$ and $V$.
\end{rem}

\begin{rem}[$N$-dependence without external fields]\label{rem-N-dependence}
    In the absence of external fields and in the physically relevant regime $\nu>2$, Griesemer and M{\o}ller \cite{griesemer-moeller-bounds-on-translation-inv-N-polaron} established stability of the Pekar-Tomasevich functional in the sense that
    \begin{align}
        \EPT^{(N,\alpha)} \, \geq \, -C\alpha^2 N \, ,
    \end{align}
    see also \cite{frank-lieb-seiringer-thomas-stability-absence}. This bound is used only in the auxiliary Lemma \ref{lem-concavity}, see also \nameref{prop-block-partition}, and improves the quadratic growth, see Remark \ref{rem-dependence-on-constant}. As a consequence, one can improve the bound the $N$-dependence of the error in Theorem \ref{thm:mainthm} to 
    \begin{equation}
		E^{(N,\alpha)} \, \geq \, \EPT^{(N,\alpha)}-CN^\frac{45}{23}\alpha^{\frac{42}{23}} \, .
	\end{equation}
    We leave the details to the interested reader. This is to be compared with the stability bound that 
    \begin{align}
        |E^{(N,\alpha)}| \, \leq \, C_\alpha N
    \end{align}
    for some constant $C_\alpha$ independent of $N$, see \cite[Theorem 3]{frank-lieb-seiringer-thomas-stability-absence}. In the regime $\nu<2$, this stability bound ceases to be true. In fact, by \cite{griesemer-moeller-bounds-on-translation-inv-N-polaron}, one has that
    \begin{align}
        -c_1N^{\frac73} \, \leq \, E^{(N,\alpha)} \, \leq \, -c_2N^{\frac73} \, .
    \end{align}
\end{rem}

\begin{rem}[Binding]
    As a consequence of Theorem \ref{thm:mainthm}, one can establish binding of the multipolaron in the case that $V=0$ and $A$ is linear, corresponding to a constant magnetic field, as long as $\nu<\nu_c(A,N)$ for some threshold $\nu_c(A,N)>2$. More precisely, one of the authors and Griesemer \cite{anapolitanos} proved the existence of $\nu_c(A,N)>2$ s.t. for all $\nu<\nu_c(A,N)$
    \begin{align}
        \min_{1\leq k\leq N}\big(\EPT^{(k,\alpha)}+\EPT^{(N-k,\alpha)}\big) \, > \, \EPT^{(N,\alpha)} \, .
    \end{align}
    Lewin \cite{lewin} proved an analogous result in the case $A=0$. As a consequence of this result and Theorem \ref{thm:mainthm}, one obtains for all $\alpha>0$ sufficiently large
    \begin{align}
        \min_{1\leq k\leq N}\big(E^{(k,\alpha)}+E^{(N-k,\alpha)}\big) \, > \, E^{(N,\alpha)} \, .
    \end{align}
\end{rem}

\begin{rem}[Vanishing of non-rescaled fields]
    As in \cite{griesemer-wellig-strong-polaron-static-fields}, one can prove that without rescaling $A\to A_\alpha$, $V\to V_\alpha$, $\lim_{\alpha\to\infty}\alpha^{-2}E^{(N,\alpha)}$ is independent of the external fields, under sufficient regularity of $A$ and $V$.
\end{rem}

We close the introduction with a brief proof sketch. In Section \ref{sec-localization} we develop the cluster localization framework; this isolates the fermionic clusters and prepares the decoupling of phonon modes. Section \ref{sec-strong-coupling} carries out the strong-coupling analysis for each cluster and derives the effective Pekar--Tomasevich energy. Section \ref{sec-proof-main} assembles the estimates to prove the main theorem. The technical appendices collect the derivation of the Pekar--Tomasevich functional and the localization and phonon-integration steps.

  \section{Cluster localization\label{sec-localization}}

 We begin by setting up the localization framework that underpins the remainder of the proof. Our goal is to localize an approximate minimizer of the Fr\"ohlich Hamiltonian in a union of balls that are far from each other. The constructed minimizer will be antisymmetric \emph{within} clusters contained in distinct balls, but not \emph{among} different balls. Due to the spatial separation of the balls, this will allow us to decompose the Hamiltonian into independent clusters. 
 \par As usual, we denote by $B_r(x)\subseteq \R^3$ the open ball of radius $r$ centered at $x$. 

\begin{lem}[Covering lemma,\cite{wellig}]\label{lem-covering-lemma}
	Let $y_1,y_2,\ldots,y_N\in\R^3$, and $R>0$. Then there are balls $B_1,B_2,\ldots,B_m$ and a partition $\{C_j\}_{j=1}^m$ of $\{1,\ldots,N\}$ such that
	\begin{enumerate}
        \item  $\ball_j$ has radius $R_j=\frac{1}{2}(3n_j-1)R\quad \forall 1\leq  j\leq m$,
		\item $\dist(\ball_i,\ball_j)\geq R$ for $i\neq j$,
		\item $\bigcup_{k\in C_j} B_R(y_k) \subseteq \ball_j$ for all $j\in\{1,\ldots,m\}$.
	\end{enumerate}  
\end{lem}
For convenience, we included the proof of Lemma \ref{lem-covering-lemma} in Appendix \ref{app-localization}.
\par In the following, we identify $\Psi\in \cH_N$  with a vector $\iota(\Psi)\in L^2(\R^{3N};\cF)$ and denote by $\supp_e(\Psi):=\supp(\iota(\Psi))$ its electronic support.

\begin{prop}[Cluster localization]\label{prop-cluster-localization}
	Let $R>0$ and $\Phi\in\mathcal{Q}_N$ be normalized. Then there exist $m$, $n_1$, $n_2$,\ldots, $n_m\in\N$ with $\sum_{j=1}^m n_j=N$, and balls $\ball_1$, $\ball_2$,$\ldots$, $\ball_m$ with the following properties.
	\begin{enumerate}
        \item $\ball_j$ has radius $R_j=\frac{1}{2}(3n_j-1)R\quad \forall 1\leq j\leq m$,
		\item $\mathrm{dist}(\ball_i,\ball_j)\geq R$ for $i\neq j$, 
		\item There exists a normalized $\Phi_0\in\Big(\bigotimes_{j=1}^m \bigwedge_{\ell=1}^{n_j}C^\infty_c(\R^3)\Big)\otimes \cF_0$ s.t.
		\begin{equation}
		\scp{\Phi}{H^{(N,\alpha)}\Phi}\geq \scp{\Phi_0}{H^{(N,\alpha)}\Phi_0}-2\pi^2N^2R^{-2}, \quad \supp_e
		(\Phi_0)\subseteq \bigtimes_{j=1}^m \ball_j^{n_j} \, .
		\end{equation}
	\end{enumerate}

\end{prop}

\begin{proof}
    As explained in the introduction, we repeat the localization procedure taken from \cite{liebloss}. Given $\chi$ a non-negative, $L^2$-normalized test function localized in $B_R(0)$, define for all $X=(x_1,\ldots,x_N)$, $Y=(y_1,\ldots,y_N)$
\begin{align}
    G(X,Y) \, := \, \sum_{\pi\in\mathcal{S}_N}\prod_{i=1}^{N}\chi(x_i-y_{\pi(i)}), \quad W_Y(X) \, :=\, \frac{G(X,Y)}{\|G(X,\cdot)\|_2} \, .
\end{align}
Let $\Phi\in \cH_N$ be normalized. With steps analogous to those in \cite[Lemma 4.1]{liebloss}, one can show that $W$ is well-defined and smooth. Employing the IMS localization formula \cite{cycon-froese-kirsch-simon,ismagilov-IMS,morgan-IMS,sigal-geometric-methods-QM-non-existence-negative-ions,simon83nondegenerateminima}, and by approximating the Dirichlet ground-state, one then argues that there exists some $Y_0\in\R^{3N}$ such that $W_{Y_0}\Phi \neq0$, and 
\begin{equation}
	\scp{W_{Y_0}\Phi}{H^{(N,\alpha)}W_{Y_0}\Phi}\leq\left(\scp{\Phi}{H^{(N,\alpha)}\Phi}+2\pi^2\frac{N^2}{R^2}\right)\scp{W_{Y_0}\Phi}{W_{Y_0}\Phi} \, .
\end{equation}
Using the Covering Lemma \ref{lem-covering-lemma}, we partition the electrons into clusters of possibly larger balls that are well-separated. Now define the joint wave-function of the separated electron clusters
\begin{align}
    \Phi_0(X) \, := \, \frac{\Big(\prod_{j=1}^m\sum_{\sigma_j\in\sym(C_j)}\prod_{\ell\in C_j} \chi(x_i-y_{\sigma_j(\ell)})\Big)\Phi(X)}{\|\Big(\prod_{j=1}^m\sum_{\sigma_j\in\sym(C_j)}\prod_{\ell\in C_j} \chi(x_i'-y_{\sigma_j(\ell)})\Big)\Phi(X')\|_{L^2_{X'}}} \, .
\end{align}
Since $H^{(N,\alpha)}$ is a local operator, and since it commutes with permutations of electrons, we can employ the fact that different electrons of different clusters are supported in disjoint sets, to obtain that 
\begin{align}
    \frac{\scp{W_{Y_0}\Phi}{H^{(N,\alpha)}W_{Y_0}\Phi}}{\|W_{Y_0}\Phi\|_2^2} \, = \, \scp{\Phi_0}{H^{(N,\alpha)}\Phi_0} \, .
\end{align}
This choice of $\Phi_0$ satisfies the needed assumptions, completing the proof.
\end{proof}

\begin{rem}[Partial statistics]
    Notice that in the proof of Prop. \ref{prop-cluster-localization}
\begin{align}
    \Big(\bigwedge_{j=1}^NL^2(\R^3)\Big)\otimes \cF \ni \frac{W_{Y_0}\Phi}{\|W_{Y_0}\Phi\|_2} \, \neq \, \Phi_0\in \Big(\bigotimes_{j=1}^m \bigwedge_{\ell=1}^{n_j}L^2(\R^3)\Big)\otimes \cF \, .
\end{align}
In particular, $\Phi_0$ is not totally antisymmetric, but antisymmetric only within clusters.
\end{rem}

\begin{rem}[Error growth with $N$]\label{rem:pay}
	In contrast to \cite{wellig}, the growth of the error in the lower bound now is of order $N^2$ and not $N$. This is due to the fact that we are accounting for fermionic statistics.
    \par Below, we will choose the optimal scaling $R=N^{-\frac{11}{30}}\alpha^{-\frac{19}{23}}$, which leads to the estimate
    \begin{equation}
		\scp{\Phi}{H^{(N,\alpha)}\Phi}\geq \scp{\Phi_0}{H^{(N,\alpha)}\Phi_0}-2\pi^2N^{\frac{82}{30}}\alpha^{\frac{38}{23}} \, .
		\end{equation}
        Recall from Rem. \ref{rem-PTscale} that $\EPT^{(N,\alpha)}=\alpha^2\EPT^{(N,1)}$. Thus the correction from localization is, indeed, of lower order. Observe that the order of magnitude of the error in $N$ coincides with the error in Theorem \ref{thm:mainthm}.
\end{rem}

In our next step, we will bound the total many-particle ground-state energy from below by the sum of the ground-state energies for less particles supported on the balls described above and the interaction between the balls. 
\par For $n\in\N$ and a Borel measurable set $M\subseteq \R^3$, denote by
\begin{equation} \label{eq-defEM}
	E^{(n,\alpha)}(M) \, := \, \inf_{\substack{\supp_e(\Phi)\subseteq M^n,\\ \|\Phi\|=1}}\scp{\Phi}{H^{(n,\alpha)}\Phi} \, .
\end{equation}
the ground-state energy of an $n$-polaron whose electrons are localized in $M$. 
\par The following proposition generalizes \cite[Lemma 3]{frank-lieb-seiringer-thomas-stability-absence}. As in \cite{wellig}, the treatment of the bipolaron is transferred to the multipolaron case. In the proof, which can be found in Appendix \ref{app-localization}, we take the fermionic statistics into account. 

\begin{prop}[Separation of clusters]\label{prop:hsplit}
	Let $\Phi_0$ be as constructed in Prop. \ref{prop-cluster-localization} and $C_j$ denote the set of electrons supported in $\ball_j$. Then $\Phi_0$ satisfies
	\begin{equation}
		\scp{\Phi_0}{H^{(N,\alpha)} \Phi_0} \; \geq \; \sum_{j=1}^mE^{(n_j,\alpha)}(\ball_j)+\sum_{i<j}\sum_{\substack{k_i\in C_i\\l_j\in C_j}}\scp{\Phi_0}{\frac{(\nu-2)\alpha}{|x_{k_i}-x_{l_j}|}\Phi_0}-\frac{8\alpha N^2}{\pi^2 R} \, .
	\end{equation}
\end{prop}

\begin{rem}[Negligibility of inter-cluster interaction]\label{rem-hsplit-coulomb-bound}
    Notice that for $x_i\in C_i$, $x_j\in C_j$, $i\neq j$, we have that $|x_i-x_j|\geq R$ by construction of $\Phi_0$. In particular, we can bound the Coulomb interaction among the different clusters by $-\frac{(\nu-2)_-\alpha N^2}{2R}$, where $x_-=\max(0,-x)$. This is the same order as the last error term in Prop. \ref{prop:hsplit}, which in our scaling choice will be negligible in the strong-coupling limit $\alpha\to\infty$.
\end{rem}

\section{Strong-coupling analysis of multipolaron clusters\label{sec-strong-coupling}}

With the cluster decomposition in place, we now carry out the strong-coupling analysis for each cluster and connect the Fr\"ohlich energy to the Pekar--Tomasevich functional.

\subsection{Pekar-Tomasevich reduction of individual clusters}

Our next goal is to justify that one can, in fact, integrate out the phonon modes to reduce the Fr\"ohlich model to the Pekar-Tomasevich model. For that, we follow the arguments in \cite{liebthomas} closely, by adapting the result to include external fields and interactions among the electrons, similarly to \cite{griesemer-wellig-strong-polaron-static-fields,wellig}. 

\par In the following, given $\Lambda>0$, we abbreviate
\begin{equation}\label{def-NE}
    \nb_\Lambda \, := \, \int_{B_\Lambda(0)} \ad(k)a(k)\dx{k}  \, .
\end{equation}
Observe, that for any $f\in L^2(B_\Lambda(0))$, we have that
\begin{equation}\label{eq-af-est}
    \|a(f)\psi\| \, \leq\, \|f\|_2\|\sqrt{\nb_\Lambda}\psi\| \, , \quad \|\ad(f)\psi\| \, \leq\, \|f\|_2\|\sqrt{\nb_\Lambda+1}\psi\| \, .
\end{equation}
For any positive integer $\enum\in\N$ and $\Lambda>0$ with $4 \enum \alpha< \pi \Lambda$, define the ultraviolet cut off Hamiltonian
\begin{equation}\label{eq-def-UVHam}
	H^{(\enum,\alpha)}_\Lambda \, := \, \sum_{j=1}^\enum\Big[ \Big(1-\frac{4n\alpha}{\pi\Lambda}\Big)D_{A_\alpha,x_j}^2+V_\alpha(x_j) +\sqrt{\alpha}\phi_\Lambda(x_j)+\sum_{k=j+1}^N\frac{\nu\alpha}{|x_j-x_k|}\Big] + \nb_\Lambda 
\end{equation}
acting on $\Big(\bigwedge_{j=1}^\enum L^2(B)\Big)\otimes \cF$, $B\subseteq \R^3$ a ball, where  
\begin{equation}
    \phi_\Lambda(x) \, := \, \frac{1}{\sqrt{2}\pi}\int_{B_{\Lambda}(0)}\frac{\dx{k}}{|k|}\big(e^{-ikx}a(k)+e^{ikx}\ad(k)\big) \, .
\end{equation}
    To account for the previous electronic localization, we introduce the space of test functions with localized electronic support
    \begin{equation}
	\cQ_{\enum}(\ball ) \, := \,\{ \Psi \in \Big(\bigwedge_{\ell=1}^\enum C_c^\infty(\R^3)\Big)\otimes\mathcal{F}_0 \text{ with } \supp_e(\Psi) \subset \ball^\enum\}.
    \end{equation}
    The following result is a straightforward extension of the result in \cite{liebthomas} to multiple interacting electrons, see \cite[Lemma 4.1]{wellig}. 
    
    \begin{lem}[UV cutoff]\label{lem-UV-cutoff}
	   For any values of $\enum\in\N$ and $\Lambda >0$ with $4 \enum \alpha< \pi \Lambda$ in the sense of quadratic forms on $\cQ_{\enum}(B)$, we have that
	   \begin{align}
		  H^{(\enum,\alpha)}\ge H^{(\enum,\alpha)}_{\Lambda} -\frac{1}{2}.\label{ineq:cutoff}
	   \end{align}
    \end{lem}

	In the next step the phonon modes are replaced by \emph{block modes}, of which only finitely many exist. More precisely, for a given $P>0$, we define
	\begin{align*}
		Q(\ell) &:=  \{k\in B_{\Lambda}(0)|k^{(i)}-\ell^{(i)}P|\le P/2\},\quad \ell\in \Z^3,\\
		\Lambda_P &:= \{\ell\in\Z^3|Q(\ell)\ne \emptyset\}.
	\end{align*}
	In every $Q(\ell)$, an arbitrary $k_\ell$ is chosen, and later optimized. The block modes are defined by
	\begin{align}\label{def-block-modes}
	   a_\ell:=\frac{1}{M_\ell}\int_{Q(\ell)}\frac{\dx{k}}{|k|}a(k),\quad M_\ell=\left(\int_{Q(\ell)}\frac{\dx{k}}{|k|^2}\right)^{1/2}.
	\end{align}
	They are well-defined normalized annihilation operators acting on the Fock space $\cF$.
    \par Let $\beta=1-\frac{4n\alpha}{\pi\Lambda}$ and define
	\begin{align}
		H_{\mathrm{block}}^{(\enum,\alpha,\beta)} \, := \, & \sum_{j=1}^\enum\left( \beta D_{A_\alpha x_j}^2+V_\alpha(x_j)+\frac{\sqrt{\alpha}}{\sqrt{2}\pi}\sum_{\ell\in\Lambda_P}M_\ell\left(e^{ik_\ell x_j}a_\ell+e^{-ik_\ell x_j}\ad_\ell\right)\right)\\
        & + \, \sum_{1\leq j<k\leq n}\frac{\nu\alpha}{|x_j-x_k|} +\beta\nb_{\mathrm{block}}, \label{def-block-mode-Hamiltonian}
	\end{align}
    where $\nb_{\mathrm{block}} := \sum_{\ell\in\Lambda_P} \ad_\ell a_\ell$. Cutting into finite many blocks allows us to repeat Pekar's argument up to a small error, as outlined after Lemma \ref{lem-pekar} in the appendix. The details are carried out in the proof of Proposition \ref{prop-block-mode-integration} in Appendix \ref{app-phonon-integration}. 
    
\begin{prop}[Block-mode reduction]\label{prop-block-partition}
    Let $\Lambda,P>0$ and $B$ be a ball of radius $r>0$. With $H^{(\enum,\alpha)}_\Lambda$ and $H_{\mathrm{block}}^{(\enum,\alpha,\beta)}$ as defined above, we then have that
    \begin{align}\label{ineq:block}
		\inf_{\substack{\Psi_n\in \cQ_{\enum}(B)\\ \|\Psi_n\|=1}} \scp{\Psi_n}{H^{(\enum,\alpha)}_\Lambda\Psi_n} \, \geq \, \inf_{\substack{\Psi_n\in \cQ_{\enum}(B)\\ \|\Psi_n\|=1}}\sup_{\{k_\ell\}}{\scp{\Psi_n}{H_{\mathrm{block}}^{(\enum,\alpha,\beta)}\Psi_n}}-\frac{6n^2P^2 r^2\Lambda\alpha}{(1-\beta)\pi} \, . 
	\end{align}
\end{prop}

    The proof, which is presented in Appendix \ref{app-localization} below, follows analogous steps to those in \cite{liebthomas}, with slight adaption due to the presence of external fields, see also \cite{wellig}. 
    \par Next, we show that the ground-state energy of the Block mode Hamiltonian can be bounded below by the Pekar-Tomasevich energy, up to an error, analogous to \cite{liebthomas}. The crucial improvement in the statement however is that we exploit the fact that the electronic support can be preserved in this bound. This insight, together with the fact that the regions, in which the clusters are localized, are well-separated, subsequently allows us to remove Assumption \eqref{eq:asswellig}, see Corollary \ref{cor-EPT-B-subadd}.

    \begin{prop}[Block mode integration]\label{prop-block-mode-integration}
	There exists a constant $C$ dependent on $A$, $V$ and $|\nu|$ such that for any $\alpha$, any $n\in\N$ and any open ball $B\subset \R^3$, we have that
	\begin{equation}\label{eq-bock-mode-integration-statement}
		\inf_{\substack{\Psi_n\in \cQ_{\enum}(B)\\\|\Psi_n\|=1}}\sup_{\{k_\ell\}}{\scp{\Psi_n}{H_{\mathrm{block}}^{(\enum,\alpha,\beta)}\Psi_n}} \, \geq \, \EPT^{(N,\alpha)}(B) - \left(2\frac{\Lambda}{P}+1\right)^3-C\frac{n^3\alpha^3}{\Lambda} \, .
	\end{equation}
    In the case $\nu>2$ and $A_k=V=0$, we obtain 
    \begin{equation}
		\inf_{\substack{\Psi_n\in \cQ_{\enum}(B)\\\|\Psi_n\|=1}}\sup_{\{k_\ell\}}{\scp{\Psi_n}{H_{\mathrm{block}}^{(\enum,\alpha,\beta)}\Psi_n}} \, \geq \, \EPT^{(N,\alpha)}(B) - \left(2\frac{\Lambda}{P}+1\right)^3-C\frac{n^2\alpha^3}{\Lambda} \, .
	\end{equation}
    \end{prop}  
    For convenience, we carried out the proof in Appendix \ref{app-phonon-integration}. Preserving the electronic support does not interfere with the arguments in \cite{liebthomas}, as we only integrate phonon degrees of freedom. Notice that $H_{\mathrm{block}}^{(\enum,\alpha,\beta)}$ does not contain the entire kinetic energy, as we sacrificed a fraction of it for the UV truncation. To effectively factor out this relative factor, which converges to 1 in the strong-coupling limit, one employs a concavity argument, as previously noted in \cite{anapolandon,griesemer-wellig-strong-polaron-static-fields,wellig}. Since the argument is not explicitly presented in these works, we outline the complete proof in Lemma \ref{lem-concavity} for the convenience of the reader.

\begin{rem}[Regime for effective Coulomb coupling]
	Observe that Proposition \ref{prop-block-mode-integration} holds for any value $\nu\in\R$. This is consistent with the fact that the assumption $\nu>0$ is not really used in the proof of \cite[Prop. 4.2]{wellig}.
\end{rem}

\subsection{Cluster synthesis}

Notice that thus far, we have kept track of the electronic support of the approximate minimizer. The electronic support is contained in a cluster of balls of comparable radii that are separated by a distance of the same order. We will now explain how this allows us to remove Assumption \eqref{eq:asswellig}. We start by explaining how the energy of disjoint clusters splits.

\begin{lem}[Energy of disjoint clusters]\label{lem-disj-cluster-energy}
    Assume that $\rho_{\psi_1}$, $\rho_{\psi_2}$, \ldots, $\rho_{\psi_m}$ have disjoint support, and assume that $\psi_j$ is an $n_j$-electron wave function. Then we have that 
    \begin{align}\label{eq-disjoint-energy-split}
        \hspace{-1ex}\HPT^{(N,\alpha)}(\psi_1\wedge\ldots\wedge \psi_m) = \sum_{j=1}^m \HPT^{(n_j,\alpha)}(\psi_j) +(\nu-2)\alpha\hspace{-0.5ex}\sum_{i <j}\int \hspace{-0.5ex}\int \hspace{-0.5ex} \frac{\rho_{\psi_i}(x) \rho_{\psi_j}(y)}{|x-y|}\dx{x} \dx{y}.
    \end{align}
\end{lem}
\begin{proof}
    Using the disjoint support of $\rho_{\psi_i}$ and $\rho_{\psi_j}$, $i\neq j$, we have that
    \begin{align}
        \rho_{\psi_1\wedge\ldots\wedge \psi_m} \, = \, \sum_{j=1}^m \rho_{\psi_j} \, .  
    \end{align}
    Observe that $\supp(\psi_j)\subseteq \supp(\rho_{\psi_j})^{n_j}$. For every local operator $\cO$ that is symmetric with respect to the coordinates $x_1,\dots,x_N$ we have that
    \begin{equation}
        \langle \psi_1 \wedge \ldots \wedge \psi_m ,\cO \psi_1 \wedge \ldots \wedge \psi_m \rangle =\langle \psi_1 \otimes \ldots \otimes \psi_m ,\cO \psi_1 \otimes \ldots \otimes \psi_m\rangle \, .
    \end{equation} 
    As a consequence, decomposing the interaction terms to the ones that are internal within a cluster and interaction terms between different clusters, we obtain
    \begin{align}
    \HPT^{(N,\alpha)}(\psi_1 \wedge \ldots\wedge \psi_m) \, &= \, \sum_{j=1}^m \HPT^{(n_j,\alpha)}(\psi_j)-\alpha \sum_{i \neq j} \int_{B_i} \int_{B_j} \frac{\rho_{\psi_i}(x) \rho_{\psi_j}(y)}{|x-y|}\dx{x}\dx{y} \\
    & \qquad + \frac{\nu \alpha }2 \sum_{i \neq j}\sum_{k \in C_i, l \in C_j} \langle \psi_i \otimes \psi_j ,\frac{1}{|x_k-x_l|} \psi_i \otimes \psi_j\rangle,
    \end{align}
    where $C_i,C_j$ denote the $i$-th and $j$-th clusters, and the factor $\frac{1}{2}$ stems from replacing  $\sum_{i < j}$ by $\sum_{i \neq j}$. Using the fact that $\rho_{\psi_i}(x)=|C_i|\int_{\R^{3(|C_i|-1)}}|\psi_i(x,X)|^2\dx{X}$, we arrive at \eqref{eq-disjoint-energy-split}.
\end{proof}

\begin{rem}[Translation-invariance and subadditivity of $\EPT$]\label{rem-EPT-subadd-TI}
    To understand the subtlety of our new auxiliary result Corollary \ref{cor-EPT-B-subadd}, let us explain the argument to prove \eqref{eq:asswellig} in the external field-free case. Let $\psi_j$ be an approximate minimizer of the Pekar-Tomasevich functional $\EPT^{(n_j,\alpha)}$ with compact support, up to error $\frac{\vep}m$. Defining the translation operator $(\tau_zf)(x_1,\ldots,x_\ell):=f(x_1+z,\ldots,x_\ell+z)$, we consider the sequence of test functions 
    \begin{align}
        \Psi_k \, := \, \tau_{k\hat{e}_1} \psi_1 \wedge \tau_{k^2\hat{e}_1} \psi_2\wedge\ldots\wedge \tau_{k^m\hat{e}_1} \psi_m \, .
    \end{align}
    For $k$ sufficiently large, we have that the supports of $\tau_{k^i\hat{e}_1} \psi_i$, $\tau_{k^j\hat{e}_1} \psi_j$, $i\neq j$, are disjoint. Using Lemma \ref{lem-disj-cluster-energy}, we then obtain
    \begin{align}
        \EPT^{(N,\alpha)} \, \leq \, \liminf_{k\to\infty} \HPT^{(N,\alpha)}(\Psi_k) \, \leq \, \sum_{j=1}^m \HPT^{(n_j,\alpha)}(\psi_j) \, \leq \, \sum_{j=1}^m \EPT^{(n_j,\alpha)}+\vep \, .
    \end{align}
\end{rem}

Since external fields break translation-invariance of $\EPT^{(n_j,\alpha)}$, we now show how to adapt the argument to our present setting. To accomplish that, we will exploit the fact that the localized clusters are well-separated.

\begin{cor}[Asymptotic subadditivity of $\EPT$]\label{cor-EPT-B-subadd}
    Let $x_+:=\max(x,0)$. Then we have that \begin{equation}
        \EPT^{(N,\alpha)} \, \leq \, \sum_{j=1}^m \EPT^{(n_j,\alpha)}(B_j) \, +\, \frac{(\nu-2)_+\alpha N^2}{2R}.\end{equation}
\end{cor}

    \begin{proof}
    Consider a normalized $n_j$-fermion state $\psi_j$ with support in $B_j$ such that $\HPT^{(n_j,\alpha)}(\psi_j)< \EPT^{(n_j,\alpha)}(B_j) + \frac{\epsilon}m $. Let $\psi_1 \wedge \ldots \wedge \psi_m$ be the normalized anti-symmetrized tensor product of $\psi_1,\dots,\psi_m$. Since the supports of $\rho_{\psi_1}, \dots, \rho_{\psi_m}$ are disjoint, Lemma \ref{lem-disj-cluster-energy} implies that
    \begin{align}
        \EPT^{(N,\alpha)} \, \leq \, \sum_{j=1}^m \HPT^{(n_j,\alpha)}(\psi_j) +\frac{(\nu-2)_+\alpha N^2}{2R} \, ,
    \end{align}
    Using the last equality together with $\HPT^{(n_j,\alpha)}(\psi_j)< \EPT^{(n_j,\alpha)}(B_j) + \frac{\epsilon}m$, and sending $\vep\to0$, we arrive at the desired result.
    \end{proof}

\section{Proof of main result\label{sec-proof-main}}

With the localization and strong-coupling estimates established, we are now in position to prove Theorem \ref{thm:mainthm}.   
\begin{proof}[Proof of Theorem \ref{thm:mainthm}]
	Let $\Phi\in\bigwedge_{j=1}^NC_c^\infty(\R^{3})\otimes\mathcal{F}_0$, $\|\Phi\|=1$ be such that
    \begin{equation}\label{eq-approx-min}
	   \scp{\Phi}{H^{(N,\alpha)}\Phi} \, \leq \, E^{(N,\alpha)} \, + \, \vep \, .
    \end{equation} 
    Employing Prop. \ref{prop-cluster-localization}, Proposition \ref{prop:hsplit} and Remark \ref{rem-hsplit-coulomb-bound}, implies that
	\begin{align}
		\scp{\Phi}{ H^{(N,\alpha)}\Phi} \, &\geq \, \sum_{j=1}^mE^{(n_j,\alpha)}(\ball_j) -\frac{2\pi^2N^2}{R^2} -\frac{8\alpha N^2}{\pi^2 R}-\frac{(\nu-2)_-\alpha N^2}{2R}
	\end{align}
	with the notation of Prop. \ref{prop-cluster-localization}. Recalling $\beta_j=1-\frac{8\enum_j \alpha}{\pi\Lambda_j}$, and collecting Lemma \ref{lem-UV-cutoff} and Propositions \ref{prop-block-partition}, and \ref{prop-block-mode-integration}, we obtain 
    \begin{align}
        E^{(\enum_j,\alpha)}(B_j) \, &\geq \, \EPT^{(\enum_j,\alpha)}(B_j)-\frac12-\Big(\frac{2\Lambda_j}{P_j}+1\Big)^3 -C\frac{n_j^3\alpha^3}{\Lambda_j} -\frac{6n_j^2P_j^2 R_j^2\Lambda_j\alpha}{(1-\beta_j)\pi} \, .
    \end{align}
    We will choose $\Lambda_j$ s.t. $\alpha/\Lambda_j\to0$ as $\alpha\to\infty$. Then let $\alpha$ be sufficiently large such that $1-\beta_j=\frac{8\enum_j \alpha}{\pi\Lambda_j}\in(0,\frac12)$. Employing Corollary \ref{cor-EPT-B-subadd} and using the fact that  for  $R_j=\frac12(3n_j-1)R\leq\frac32n_jR$, we obtain that
    \begin{align}
        \scp{\Phi}{ H^{(N,\alpha)}\Phi} \, &\geq \, \EPT^{(N,\alpha)}-\frac{|\nu-2|\alpha N^2}{2R}-\frac{2\pi^2N^2}{R^2} -\frac{8\alpha N^2}{\pi^2 R}-\frac{m}2\\
        &\qquad -\sum_{j=1}^m\Big(\frac{2\Lambda_j}{P_j}+1\Big)^3-C\sum_{j=1}^m\frac{n_j^3\alpha^3}{\Lambda_j}-\frac{27}{16}R^2\sum_{j=1}^mn_j^3 P_j^2 \Lambda_j^2 \, .
    \end{align}
    Due to $\sum_{j=1}^mn_j^q \leq N^q$ for $q\geq1$, the optimal choice
    \begin{align}
        \Lambda_j \, = \, n_j^{\frac{19}{30}}\alpha^{\frac{27}{23}} , \quad P_j\, = \, n_j^{-\frac25}\alpha^{\frac{13}{23}} , \quad R \, = \, N^{-\frac{11}{30}}\alpha^{-\frac{19}{23}} 
    \end{align}
    in terms of power of $\alpha$ and $N$ yields for sufficiently large $\alpha>0$
    \begin{align}
        E^{(N,\alpha)} \, \geq \, \EPT^{(N,\alpha)}-CN^{\frac{82}{30}}\alpha^{\frac{42}{23}}
    \end{align}
    for some constant $C$ dependent on $A$, $V$, and $|\nu|$, as clarified in Remark \ref{rem-dependence-on-constant}; this dependence arises from the bound established in Lemma \ref{lem-concavity} in the appendix. This completes the proof of our main Theorem \ref{thm:mainthm}.
\end{proof}

\appendix

\section{Derivation of the Pekar-Tomasevich functional\label{app-PT}}

\subsection{Pekar's argument}

We now present the calculation leading to the reduction to $\EPT^{(N,\alpha)}$, see \eqref{eq:ptdef}. 
\begin{lem}[Calculation of $\HPT$]\label{lem-pekar}
	We have that
	\begin{align}
		\HPT^{(N,\alpha)}(\Psi_N) \, &= \, \sum_{j=1}^N\scp{\Psi_N}{\Big(D_{A_\alpha,x_j}^2+V_\alpha(x_j)+\sum_{k=j+1}^N\frac{\nu\alpha }{|x_j-x_k|}\Big)\Psi_N}\\
        &\qquad -\alpha \int\frac{\rho_{\Psi_N}(x)\rho_{\Psi_N}(y)}{|x-y|}\dx{x}\dx{y} \, .
	\end{align}
\end{lem}
 
 \begin{proof}
 	Notice that for $\|\eta\|_{\mathcal{F}}=\|\Psi_N\|_{\bigwedge_{j=1}^N L^2(\R^3)}=1$
\begin{align}
	\MoveEqLeft\scp{\Psi_N\otimes\eta}{H^{(N,\alpha)}\Psi_N\otimes\eta}\\
    &= \, 
	\sum_{j=1}^N\scp{\Psi_N}{\Big(D_{A_\alpha,x_j}^2+V_\alpha(x_j)+\sum_{k=j+1}^N\frac{\nu\alpha }{|x_j-x_k|}\Big)\Psi_N}\\
	&\qquad +\int\scp{\eta}{\left(\ad(k)a(k)+\frac{\sqrt{\alpha}}{\sqrt{2}\pi|k|}\left(a(k)\overline{\widehat{\rho_\psi}}(k)+\ad(k)\widehat{\rho_\psi}(k)\right)\right)\eta}  \dx{k} \, .
\end{align}
Here $\widehat{\rho_{\Psi_N}}$ denotes the Fourier transform of the electron density $\rho_{\Psi_N}$ defined in \eqref{def-density}. By completing the square, we arrive at
\begin{align}
	\scp{\Psi_N\otimes\eta}{H^{(N,\alpha)}\Psi_N\otimes\eta} \, =&\, \sum_{j=1}^N\scp{\Psi_N}{\Big(D_{A_\alpha,x_j}^2+V_\alpha(x_j)+\sum_{k=j+1}^N\frac{\nu\alpha }{|x_j-x_k|}\Big)\Psi_N} \\
	&+ \, \int\|\left(a(k)+\frac{\sqrt{\alpha}\widehat{\rho_\psi}(k)}{\sqrt{2}\pi|k|}\right)\eta\|^2  \dx{k}-\frac{\alpha}{2\pi^2}\int  \frac{|\widehat{\rho_\psi}|^2}{|k|^2} \dx{k} \, .
\end{align}
In order to minimize this expression, we choose a sequence in $\mathcal{F}_0$ approximating a coherent state such that the second to last term vanishes. By applying Plancherel and the fact that the inverse Fourier transform of $\frac{1}{|k|^2}$ satisfies
\begin{equation}
	\widecheck{|k|^{-2}}(x) \, = \, (2\pi)^{-3}\int\frac{\e{\i kx}}{|k|^2} \dx{k} \, = \, \frac{2\pi^2}{|x|},\label{eq:k-2ft}
\end{equation}
we then obtain
\begin{align}
	\HPT^{(N,\alpha)}(\Psi_N) \, = &\, \sum_{j=1}^N\scp{\Psi_N}{\Big(D_{A_\alpha,x_j}^2+V_\alpha(x_j)+\sum_{k=j+1}^N\frac{\nu\alpha }{|x_j-x_k|}\Big)\Psi_N} \\
	&- \,\alpha \int  \frac{\rho_{\Psi_N}(x)\rho_{\Psi_N}(y)}{|x-y|}\dx{x}\dx{y} \, .
\end{align}
This finishes the proof.
\end{proof} 

\subsection{Sketch of rigorous version of Pekar's argument}

As we establish in the proof of \nameref{prop-block-mode-integration}, Pekar's argument can be rigorously applied to the block mode Hamiltonian $H^{(\enum,\alpha,\beta)}_{\mathrm{block}}$. To explain the basic idea, let us consider the simplest case $\enum=1,\beta=1$, $A_k=V=0$
\begin{align}
    H_\alpha \, := \, -\Delta+\frac{\sqrt{\alpha}}{\sqrt{2}\pi}\sum_{\ell\in\Lambda_P}M_\ell\Big(a_\ell e^{ik_\ell\cdot x}+\ad_\ell e^{-ik_\ell\cdot x}\Big)+\sum_{\ell\in\Lambda_P}\ad_\ell a_\ell \, .
\end{align}
Let $\cF_{\Lambda_P}$ denote the Fock space generated by the CCR algebra generated by $a_\ell^{\#}$, i.e.,
\begin{align}
    \cF_{\Lambda_P} \, = \, \overline{\Span\{\prod_{\ell\in\Lambda_P}(\ad_\ell)^{n_\ell}\Omega\mid n_\ell\in \N \, \forall \ell\in\Lambda_P\}}^{\cF} \, .
\end{align}
We define the \emph{coherent states}
\begin{align}
    \eta_z \, := \, \prod_{\ell\in\Lambda_P}e^{z_\ell\ad_\ell-\overline{z}_\ell a_\ell}\Omega \, .
\end{align}
Abbreviating $\dx{z}=\prod_{\ell\in\Lambda_P}\frac{\dx{x_\ell}\dx{y_\ell}}\pi$, $z_\ell=x_\ell+i y_\ell$, we can expand any $\psi\in L^2(\R^3)\otimes \cF_{\Lambda_P}$
\begin{align}
    \psi \, = \, \int \psi_z \eta_z \dx{z}, \quad \psi_z \, := \, \scp{\eta_z}{\psi} \, .
\end{align}
In particular, we have decomposed $\psi$ into an integral over product states, rather than a single product state. Using the fact that
\begin{align}
    \scp{\eta_z}{\ad_\ell a_\ell \eta_z} \, = \, |z_\ell|^2-1 \, ,
\end{align}
one can employ the analogous steps of Lemma \ref{lem-pekar} to obtain that
\begin{align}
    \scp{\psi}{H_\alpha \psi} \, \geq \, \EPT^{(1,\alpha)}-\sum_{\ell \in\Lambda_P}1 \, = \, \EPT^{(1,\alpha)}-|\Lambda_P| \, .
\end{align}
Finally, we have the freedom to choose the parameters $\Lambda,P$ s.t. $|\Lambda_P|\ll \alpha^2\sim E^{(1,\alpha)}$. For more details, we refer to Appendix \ref{app-phonon-integration}.

\section{Localization arguments \label{app-localization}}

\begin{proof}[Proof of Lemma \ref{lem-covering-lemma}]
	Assume the statement holds true for $N$. Consider an additional point $y_{N+1}\in\R^3$. If $\dist(B_R(y_{N+1}),\ball_j)\geq R$ for all $j\in\{1,\ldots,m\}$, define $B_{m+1}:=B_R(y_{N+1})$ and $C_{m+1}:=\{N+1\}$.
	\par If $\dist(B_R(y_{N+1}),B_{j_1})<R$ for some $j_1\in \{1,\ldots,m\}$, there is a ball $\widetilde{B}_{j_1}\supseteq B_R(y_{N+1})\cup B_{j_1}$ with radius $\frac12[3(|C_{j_1}|+1)-1]R$, and we define $\widetilde{C}_{j_1}:=C_{j_1}\cup\{y_{N+1}\}$. If there is $B_{j_2}$, $j_2\in\{1,\ldots,N\}\setminus\{j_1\}$, such that $\dist(\widetilde{B}_{j_1},B_{j_2}) <R$, there is a ball $\widetilde{B}_{j_2}\supseteq \widetilde{B}_{j_1}\cup B_{j_2}$ with radius $\frac12[3(|\widetilde{C}_{j_1}|+|C_{j_2}|)-1]R$, and we define $\widetilde{C}_{j_2}:=\widetilde{C}_{j_1}\cup C_2$. By repeating this procedure, we conclude the proof. 
\end{proof}

\begin{proof}[Proof of Prop. \ref{prop:hsplit}]
  As in \cite{wellig}, we start by subdividing $\R^3$ into the 'area of influence' of the single multipolarons. More precisely, we define
	\begin{equation}
		S_j\;:=\;\{x\in\R^3|\;\mathrm{dist}(\ball_j,x)<\mathrm{dist}(\ball_k,x)\;\forall k\neq j\}\nonumber
	\end{equation}
	and find $S_j\cap S_k=\emptyset$ for any $j\neq k$. For $\mathrm{dist}(\ball_j,\ball_k)>0$ if $j\neq k$, we have $\ball_j\subseteq S_j$ as well as $\overline{\bigcup_{j=1}^mS_j}=\R^3$. 
	\par Next, we decompose $H^{(N,\alpha)}$ with respect to this spatial partition. We write
	\begin{align}
		H^{(N,\alpha)}=&\sum_{j=1}^m\Bigg(\sum_{\ell\in C_j}\Big(D_{A_\alpha,x_\ell}^2+V_\alpha(x_\ell)+\sqrt{\alpha}\phi(x_\ell)\Big)+\sum_{r,s\in C_j}\frac{\nu\alpha}{|x_r-x_s|}\Bigg)\nonumber\\
		&+\nb+\sum_{j<k}\sum_{\substack{r\in C_j,\\ s\in C_k}}\frac{\nu\alpha}{|x_{r}-x_{s}|} \label{eq-HNAcluster} \, ,
	\end{align} 
	where $C_j$ denotes the set of electrons supported in $\ball_j$. Then we define the operator-valued distributions
	\begin{equation}
		\ca(x) \, := \, \frac{1}{(2\pi)^{3/2}}\int \e{\i kx}a(k) \dx{k},\quad \cad(x):=\frac{1}{(2\pi)^{3/2}}\int\e{-\i kx} \ad(k)\dx{k}
	\end{equation}
	on $\mathcal{F}$. Plancherel's theorem implies that
	\begin{equation}
		\phi(x) \, = \, \frac{1}{\pi^{3/2}}\int \frac{\ca(y)+\cad(y)}{|x-y|^2} \dx{y}, \quad \Hph=\int \cad(y)\ca(y) \dx{y} \, . \label{eq-phonon-terms-0}
	\end{equation}
    We now want to localize the electron positions in the phonon terms. For that, observe that
    \begin{align}
        \MoveEqLeft\nb+\sqrt{\alpha}\sum_{j=1}^N\phi(x_j)  =\\
     \sum_{j=1}^m\int_{S_j}&\Big(\cad(y)\ca(y)+\frac{\sqrt{\alpha}}{\pi^{3/2}}\sum_{\ell\in C_j}\frac{\ca(y)+\cad(y)}{|x_\ell-y|^2} + (\ca(y)+\cad(y))g_j(y,X_{C_j^c})\Big)\dx{y} \, , \label{eq-phonon-localized}
    \end{align}
    where
	\begin{equation}
		g_j(y,X_{C_j^c})\;:=\;\frac{\sqrt{\alpha}}{\pi^{3/2}}\sum_{k\neq j}\sum_{\ell\in C_k}\frac{1}{|x_\ell-y|^2}\mathds{1}_{S_j}(y) , \quad X_{C_j^c}=(x_k)_{\substack{1\leq k\leq N,\\ k\notin C_j}} \, . \label{eq-loc-phon-int-0}
	\end{equation}
    The first term in \eqref{eq-phonon-localized} corresponds to the localized phonon field energy, and the second to the localized electron-phonon coupling. The last term in \eqref{eq-phonon-localized} corresponds to the interactions with all other clusters.
    To absorb the latter, we conjugate the Hamiltonian with the unitary Weyl transform 
    \begin{align}
        \cW_j \, := \, \cW[g_j(\cdot,X_{C_j^c})] \, ,
    \end{align}
    where for $f\in L^2(\R^3)$ we have that
    \begin{equation}
        \cW[f] \, := \, \exp\Big(\int\big(f(x)\cad(x)-\bar{f}(x)\ca(x)\big)\dx{x} \Big) \, .
    \end{equation}
    Recall that
    \begin{align}\label{eq-Weyltrans}
        \cW^\dagger[f]\ca(x)\cW[f] \, = \, \ca(x)+f(x), \quad \cW^\dagger[f]\cad(x)\cW[f] \, = \, \cad(x)+\bar{f}(x) \, .
    \end{align}
    This, in turn leads to scalar corrections, which we will show to be of lower order below. More precisely, we define the restricted field and coupling operators
    \begin{align}
        \nb_j \, &:= \,\int_{S_j}\cad(y)\ca(y) \dx{y}\, ,\\
        \phi_j(x) \, &:= \, \int_{S_j}\frac{\ca(y)+\cad(y)}{|x-y|^2} \dx{y} \, , \quad x \in S_j \, ,
    \end{align}
    and the error terms
    \begin{align} \label{def-f12}
        F_1(X) \, := \, \sum_{j=1}^m\|g_j(\cdot,X_{C_j^c})\|_2^2,\quad F_2(X)\, := \,\frac{2\sqrt{\alpha}}{\pi^{3/2}}\sum_{j=1}^m\sum_{\ell\in C_j}\int_{S_j}\frac{g_j(y,X_{C_j^c})}{|x_\ell-y|^2} \dx{y} \, ,
    \end{align}
    where $X=(x_1,\ldots, x_N)$. Using \eqref{eq-Weyltrans}, a lengthy but elementary calculation leads to
    \begin{align}\label{eq-WjstarWj}
        \nb+\sqrt{\alpha}\sum_{j=1}^N\phi(x_j) \, = \, \sum_{j=1}^m\cW_j^\dagger\Big(\nb_j+\sqrt{\alpha}\sum_{\ell\in C_j}\phi_j(x_\ell)\Big)\cW_j-(F_1+F_2) \, .
    \end{align}
    Moreover, since $g_j(.,X_{C_j^c})$ commutes with all operators acting only on coordinates of the cluster $C_j$, we have that
    \begin{align}
        \MoveEqLeft\sum_{\ell\in C_j}\big(D_{A_\alpha,x_\ell}^2+V_\alpha(x_\ell)\big)+ \sum_{\substack{r,s\in C_j\\ r<s}}\frac{\nu\alpha}{|x_r-x_s|} \\ \label{eq-Unitinv}
         &= \, \cW_j^\dagger   \Big[\sum_{\ell\in C_j}\big(D_{A_\alpha,x_\ell}^2+V_\alpha(x_\ell)\big)+ \sum_{\substack{r,s\in C_j\\ r<s}}\frac{\nu\alpha}{|x_r-x_s|}\Big]\cW_j \, .
    \end{align}
	Define
	\begin{align}
		H_j^{(n_j,\alpha)} \; :=& \; \sum_{\ell\in C_j}\big(D_{A_\alpha,x_\ell}^2+V_\alpha(x_\ell)+\sqrt{\alpha}\phi_j(x_\ell)\big) + \nb_j + \sum_{\substack{r,s\in C_j\\ r<s}}\frac{\nu\alpha}{|x_r-x_s|}.
		\label{eq:kidef}
	\end{align}
	This term corresponds to the energy of the electrons supported within $\ball_j$ and the phonons which are closer to this ball than to any other ball, including the electron-phonon interaction between these. Using \eqref{eq-HNAcluster},\eqref{eq-WjstarWj},\eqref{eq-Unitinv} and \eqref{eq:kidef} we obtain
	\begin{align}
		\MoveEqLeft\scp{\Phi_0}{H^{(N,\alpha)} \Phi_0}  = \, \sum_{j=1}^m\scp{\Phi_0}{\cW_j^\dagger H_j^{(n_j,\alpha)}\cW_j\Phi_0} \\ \label{eq-Phi0HPhi0}
       &+\sum_{j<k}\sum_{\substack{r\in C_j\\s\in C_k}}\scp{\Phi_0}{\frac{\nu\alpha}{|x_r-x_s|}\Phi_0}-\scp{\Phi_0}{(F_1+F_2)\Phi_0} \, .
	\end{align}
    By unitarity $\cW_j=\cW[g_j(\cdot,X_{C_j^c})]$ and by its locality in the electronic coordinates $X=(x_1,\ldots,x_N)$, we find, with the help of \eqref{eq-defEM}, that
    \begin{align}\label{ineq-Weyl}
        \scp{\Phi_0}{\cW_j^\dagger H_j^{(n_j,\alpha)}\cW_j\Phi_0} \, \geq \, E^{(n_j,\alpha)}(B_j) \, .
    \end{align}
    By Lemma \ref{lem:fibd} below, we also have that
    \begin{align}\label{ineq-LemmaDavid}
        \scp{\Phi_0}{F_1\Phi_0} \, \leq \, \frac{8\alpha N^2}{\pi^2R}, \quad \scp{\Phi_0}{F_2\Phi_0}\, \leq \, 2\alpha\sum_{j<k}\sum_{\substack{r\in C_j\\ s\in C_k}}\scp{\Phi_0}{\frac{1}{|x_{r}-x_{s}|}\Phi_0}.
    \end{align}
  Using \eqref{eq-Phi0HPhi0}, \eqref{ineq-Weyl}, and \eqref{ineq-LemmaDavid} we thus conclude the proof of Proposition \ref{prop:hsplit}.
    \end{proof}

    The following result is taken directly from \cite[Lemma 3.4]{wellig}. 

    \begin{lem}[Inter-cluster interaction bounds]\label{lem:fibd}
	   Let $\Phi_0$ be as constructed in Prop. \ref{prop-cluster-localization}, and recall $F_1$ and $F_2$ from \eqref{def-f12}. Then we have
	   \begin{enumerate}
		\item $\scp{\Phi_0}{F_1\Phi_0}\;\leq\; \frac{8\alpha N^2}{\pi^2R}$,
		\item $\displaystyle\scp{\Phi_0}{F_2\Phi_0}\;\leq\;2\alpha\sum_{j<k}\sum_{\substack{r\in C_j\\ s\in C_k}}\scp{\Phi_0}{\frac{1}{|x_{r}-x_{s}|}\Phi_0}.$
	   \end{enumerate}
    \end{lem}
\begin{proof}
For the proof we refer to \cite[Lemma 3.4]{wellig}. The proof of part (ii) relies on the known identity
\begin{equation}
\int_{\R^3}\frac{1}{|x|^2|x-y|^2}\dx{x} \, = \, \frac{\pi^3}{|y|} \, ,
\end{equation}
which we want to want to explain in the following. Due to the invariance of the Lebesgue measure under rotations, we may assume that $y=(0,0,|y|)$. Using spherical coordinates, we find, after integrating with respect to the angle variables, that
\begin{equation}
\int_{\R^3}\frac{1}{|x|^2|x-y|^2}\dx{x} \, = \, \frac{2\pi}{|y|}\int_0^\infty \frac{1}{r} \log\left(\frac{r+|y|}{|r-|y||}\right)\dx{r} \, = \, \frac{2\pi}{|y|}\int_0^\infty \frac{1}{t} \log\left(\frac{t+1}{|t-1|}\right)\dx{t} \, ,
\end{equation}
where in the last step, we substituted $t=\frac{r}{|y|}$. To compute the last integral, observe that substituting $s=1/t$ implies that
\begin{equation}
    \int_1^\infty \frac{1}{t} \log\left(\frac{t+1}{t-1}\right)\dx{t} \, = \, \int_0^1 \frac1s\log\left(\frac{s+1}{1-s}\right)\dx{s} \, .
\end{equation}
 In particular, we have that
\begin{align}
    \int_{\R^3}\frac{1}{|x|^2|x-y|^2}\dx{x} \, = \, \frac{4\pi}{|y|}\int_0^1 \frac{1}{t} \log\left(\frac{t+1}{1-t}\right)\dx{t} \, .
\end{align}
By differentiating and integrating, we find the identity 
\begin{align}
    \log\left(\frac{t+1}{1-t}\right) \, = \, 2 \sum_{n=0}^\infty \frac{t^{2n+1}}{2n+1} \quad \forall t \in [0,1) \, .
\end{align}
Plugging in this series, integrating term by term and using the identity
\begin{align}
    \sum_{n=0}^\infty\frac1{(2n+1)^2} \, = \, \sum_{n=1}^\infty\frac1{n^2}-\sum_{n=1}^\infty\frac1{(2n)^2} \, = \, \frac34 \sum_{n=1}^\infty\frac1{n^2} \, = \, \frac{\pi^2}8 \, ,
\end{align}
the monotone convergence implies the desired result.
\end{proof}

\section{Phonon integration \label{app-phonon-integration}}
We now provide the proofs of \cite{liebthomas} with their proper adaption to the multipolaron setting, see also \cite{wellig}. The reason, for which we present these proofs, is to guarantee that the localized electronic support inside of a given ball can, in fact, be preserved.

\begin{proof}[Proof of \nameref{prop-block-partition}]
    We start by shifting the center $x_0$ of $B$ to the origin. To that end, we introduce the total momentum
    \begin{equation}
        \mathbf{P} \, := \, -i\nabla_x+\int k\ad(k) a(k)\dx{k} \, .
    \end{equation}
    Conjugating $H^{(\enum,\alpha)}_\Lambda$ with $\exp(-ix_0\cdot \mathbf{P})$ then corresponds to replacing the external potentials by 
    \begin{equation}
        V_\alpha(x)\to  V_\alpha(x+x_0), \quad A_\alpha(x)\to A_\alpha(x+x_0) \, ,
    \end{equation}
    and the center of $B$ by the origin. Since the ground-state energy is invariant under conjugation with unitary transformations, it suffices to show the statement for the shifted external potentials. 
    \par Now let $1\le j\le n$. We define
    \begin{align}
        \nb_{Q(\ell)} \, := \, \int_{Q(\ell)}\ad(k)a(k)\dx{k} \, .
    \end{align}
    Then in the sense of quadratic forms 
	\begin{align}
		\MoveEqLeft\frac{1-\beta}{\enum} \nb_{Q(\ell)} +\frac{\sqrt{\alpha}}{\sqrt{2} \pi}\int_{Q(\ell)} \frac{1}{|k|}\left( (e^{ikx_j}-e^{ik_\ell x_j})a(k)+(e^{-ikx_j}-e^{-ik_\ell x_j})\ad(k)\right)\dx{k} \nonumber\\ 
		& \, \geq \, -\frac{\enum\alpha}{2\pi^2(1-\beta)}\int_{Q(\ell)}\frac{|e^{ikx_j}-e^{ik_\ell x_j}|^2}{|k|^2} \dx{k},
	\end{align}
	which follows from completion of squares. For any $k\in Q(\ell)$, $\ell\in\Lambda_P$ and $|x_j|< r$, $1\leq j\leq \enum$, we have that
	\begin{align}\label{ineq:eikneik}
		|e^{ikx_j}-e^{ik_\ell x_j}|^2 \, \leq \, |k-k_\ell|^2|x_j|^2\, \leq \, 3P^2 r^2  \, .
	\end{align}
    Moreover, we have that $\int_{B_\Lambda(0)}\frac{\dx{k}}{|k|^2}=4\pi\Lambda$. As a consequence, we find that
    \begin{align}
        \MoveEqLeft\sum_{j=1}^\enum\sum_{\ell\in\Lambda_P} \Big(\frac{1-\beta}{\enum} \nb_{Q(\ell)} +\frac{\sqrt{\alpha}}{\sqrt{2} \pi}\int_{Q(\ell)} \left( (e^{ikx_j}-e^{ik_\ell x_j})a(k)+(e^{-ikx_j}-e^{-ik_\ell x_j})\ad(k)\right)\frac{\dx{k}}{|k|} \Big) \nonumber\\
		& \, \geq \, -\frac{6\enum^2\alpha \Lambda P^2r^2}{\pi(1-\beta)}  \, .\label{ineq:knquaderg}
    \end{align}
    Recalling \eqref{def-block-modes}, observe that due to $\langle\Psi,\ad_\ell a_\ell \Psi \rangle= \|a_\ell \Psi\|^2$ and 
    \begin{align}
        \|a_\ell \Psi\| \, &= \, \frac1{M_\ell}\|\int_{Q(\ell)}\frac{1}{|k|}a(k)\dx{k}\Psi\| \, \leq \, \frac1{M_\ell }\int_{Q(\ell)}\frac{1}{|k|} \|a(k)\Psi\|\dx{k} \\& \leq\, \Big(\int_{Q(\ell)}\|a(k)\Psi\|^2\dx{k}\Big)^{\frac12} \, = \, \langle \Psi, \nb_{Q(\ell)}\Psi\rangle^{\frac12} ,
    \end{align}
    we have that 
    \begin{align}\label{eq-nbblock-nb-Lambda}
        \nb_{\mathrm{block}} \, = \, \sum_{\ell \in \Lambda_P} \ad_\ell a_\ell \, \leq \, \sum_{\ell \in \Lambda_P} \nb_{Q(\ell)} \, = \, \nb_{\Lambda} \, . 
    \end{align}
    We rewrite 
    \begin{align}
        \nb_{\Lambda} \, = \, \beta \nb_{\Lambda} + (1-\beta)\nb_{\Lambda} \, = \, \beta\nb_{\Lambda} +\sum_{j=1}^n\sum_{\ell\in \Lambda_P}\frac{1-\beta}n  \nb_{Q(\ell)}
    \end{align}
    and with the help of \eqref{eq-def-UVHam} and \eqref{def-block-mode-Hamiltonian} we find
    \begin{align}
        \MoveEqLeft H^{(\enum,\alpha)}_{\Lambda} \, =   H^{(\enum,\alpha,\beta)}_{\mathrm{block}}+\beta(\nb_{\Lambda}-\nb_{\mathrm{block}})+\sum_{j=1}^{\enum}\sum_{\ell\in\Lambda_P}\Big(\frac{1-\beta}{\enum}\nb_{Q(\ell)}\\ 
        & \quad +\frac{\sqrt{\alpha}}{\sqrt{2} \pi}\int_{Q(\ell)} \frac{\dx{k}}{|k|}\left( (e^{ikx_j}-e^{ik_\ell x_j})a(k)+(e^{-ikx_j}-e^{-ik_\ell x_j})\ad(k)\right)\Big),
    \end{align}
    where both, $H^{(\enum,\alpha)}_{\Lambda}$ and $H^{(\enum,\alpha,\beta)}_{\mathrm{block}}$, contain the shifted fields $V_\alpha$, $A_\alpha$. Since $k_\ell$ was chosen arbitrarily, we employ \eqref{ineq:knquaderg} and \eqref{eq-nbblock-nb-Lambda} to bound the error terms. Modulo translating $H^{(\enum,\alpha)}_{\Lambda}$ and $H^{(\enum,\alpha,\beta)}_{\mathrm{block}}$ back, we thus obtain that
    \begin{align}
		\inf_{\substack{\Psi_n\in \cQ_{\enum}(B)\\ \|\Psi_n\|=1}} \scp{\Psi_n}{H^{(\enum,\alpha)}_\Lambda\Psi_n} \, \geq \, \inf_{\substack{\Psi_n\in \cQ_{\enum}(B)\\ \|\Psi_n\|=1}}\sup_{\{k_\ell\}}{\scp{\Psi_n}{H_{\mathrm{block}}^{(\enum,\alpha,\beta)}\Psi_n}}-\frac{6n^2P^2 r^2\Lambda\alpha}{(1-\beta)\pi} \, . 
	\end{align}    
    This concludes the proof.
\end{proof}

Before we prove \nameref{prop-block-mode-integration}, it is useful to introduce some notation. Recalling definitions \eqref{def-density} of $\rho_\Psi$ and \eqref{def-EPT} of $\EPT$, we introduce
\begin{align}
    \EPT^{(\enum,\alpha,\lambda)}(B) \, &:= \, \inf_{\substack{\Psi_\enum\in \cQ_{\enum}(B)\\\|\Psi_\enum\|=1}}\Bigg( \scp{\Psi_\enum}{\Big(\sum_{j=1}^\enum\big(D_{A_\alpha,x_j}^2+ \lambda V_\alpha(x_j)\big)+ \sum_{k=j+1}^N\frac{\lambda\nu\alpha}{|x_j-x_k|}\Big)\Psi_n}\\ \label{def-EPT-concave}
    &\qquad -\alpha\lambda^2 \int_{B^2}\frac{\rho_{\Psi_n}(x)\rho_{\Psi_n}(y)}{|x-y|} \dx{x}\dx{y} \Bigg) \, .
\end{align}
Then we have the following result.

\begin{lem}[Concavity bound]\label{lem-concavity}
    Assume that $\beta=1-\frac{4n\alpha}{\pi\Lambda}\in(\frac12,1)$. There exists a constant $C>0$ dependent on $A$, $V$, and $|\nu|$ s.t.
    \begin{align}
        \beta \EPT^{(n,\alpha,\beta^{-1})}(B) \, \geq \, \EPT^{(N,\alpha)}(B)-C\frac{n^3\alpha^3}{\Lambda} \, .
    \end{align}
    In the absence of $A$ and $V$, this bound can be improved to
    \begin{align}
        \beta \EPT^{(n,\alpha,\beta^{-1})}(B) \, \geq \, \EPT^{(N,\alpha)}(B)-C\frac{n^2\alpha^3}{\Lambda} \, .
    \end{align}
\end{lem}

\begin{proof}
    Building upon ideas in \cite{wellig}, we reduce $\EPT^{(\enum,\alpha,\beta^{-1})}(B)$ to $\EPT^{(\enum,\alpha)}(B)$ by exploiting concavity of $\lambda\mapsto\EPT^{(n,\alpha,\lambda)}(B)$. More precisely, 
    as an infimum over concave functions, $\lambda\mapsto \EPT^{(\enum,\alpha,\lambda)}(B)$ itself is concave. In particular, since $\beta^{-1}\in(1,2)$, we obtain that 
    \begin{align}\label{eq-EPT-concave}
        \EPT^{(\enum,\alpha,\beta^{-1})}(B) \, \geq \, (2-\beta^{-1})\EPT^{(\enum,\alpha,1)}(B)+(\beta^{-1}-1)\EPT^{(\enum,\alpha,2)}(B) \, ,
    \end{align}
    and we recall that $\EPT^{(\enum,\alpha,1)}(B)=\EPT^{(\enum,\alpha)}(B)$. 
    \par To complete the proof, it suffices to establish an upper bound on $\EPT^{(\enum,\alpha,\lambda)}$. A scaling argument analogous to Remark \ref{rem-PTscale} yields that
    \begin{align}
        \EPT^{(n,\alpha,\lambda)} (B) \, = \, \alpha^2\EPT^{(n,1,\lambda)}(\alpha^{-1}B) \, .
    \end{align}
    Hardy's inequality and the diamagnetic inequality \cite[Theorem 7.21]{llbook}, together with the infinitesimal form-boundedness of $V$ w.r.t. $-\Delta$, and $A_k\in L^2_{\mathrm{loc}}(\R^3)$ imply the general bound 
    \begin{align}
        \min_{\lambda=1,2}\EPT^{(n,1,\lambda)}(\alpha^{-1}B) \, \geq \, \min_{\lambda=1,2}\EPT^{(n,1,\lambda)}(\R^3) \, \geq \, - Cn^2
    \end{align}
    for some constant $C>0$ dependent on $A$,$V$ and $|\nu|$. 
    In the case $\nu>2$ and in the absence of external fields $A$, $V$, \cite[Theorem 3]{frank-lieb-seiringer-thomas-stability-absence} implies
    \begin{align}\label{eq-pekar-polaron-stability}
        \min_{\lambda=1,2}\EPT^{(n,1,\lambda)}(\alpha^{-1}B) \, \geq \, \min_{\lambda=1,2}\EPT^{(n,1,\lambda)}(\R^3) \, \geq \, - Cn
    \end{align}
    for some $C=C(\nu)>0$, where $\lim_{\nu\to2^+}C(\nu)\to\infty$. Their result holds independently of statistics of particles. The case $\nu=2$ without external fields is covered in \cite{griesemer-moeller-bounds-on-translation-inv-N-polaron} and establishes the stability bound \eqref{eq-pekar-polaron-stability} in the fermionic case. Together with \eqref{eq-EPT-concave}, we conclude the proof.
\end{proof}

\begin{proof}[Proof of \nameref{prop-block-mode-integration}]
    We will prove that for all normalized $\psi\in \cQ_{\enum}(B)$, we have that
	\begin{align}\label{eq:pekar}
		\sup_{\{k_\ell\}}{\scp{\psi}{H_{\mathrm{block}}^{(\enum,\alpha,\beta)}\psi}} \, \geq \, \beta\EPT^{(n,\alpha,\beta^{-1})}(B) - \beta |\Lambda_P|.
	\end{align}
    The statement then follows from Lemma \ref{lem-concavity} and the facts that $|\Lambda_P|\le\left(2\frac{\Lambda}{P}+1\right)^3$ and that $\beta\leq1$. 
	\par To prove \eqref{eq:pekar}, the block operators $a_\ell$ are replaced by complex numbers $z_\ell$ using coherent states. The closed subspace $M:=\text{span}\{\chi_{Q(\ell)}|.|^{-1}|\ell\in\Lambda_P\}\subset L^2(\R^3)$ generates the symmetric Fock space $\cF (M)$, i.e. the Fock space that is constructed by the block operators $\ad_\ell$, $\ell\in\Lambda_P$. Since $M$ is a closed subspace
	\begin{align*}
		\cF=\cF(M\oplus M^{\perp})\cong \cF(M)\otimes\cF(M^{\perp}).
	\end{align*}
	Suppose $z=(z_\ell)_{\ell\in\Lambda_P}$, $z_\ell\in\C$. Then we define normalized coherent states $\eta_z\in \cF (M)$ by
	\begin{align}
		\eta_z:=\prod_{\ell\in\Lambda_P}e^{z_\ell\ad_\ell-\overline{z}_\ell a_\ell}\Omega,\label{def:cohrentstate}
	\end{align}
	where $\Omega\in\cF (M)$ denotes the normalized vacuum. Coherent states have the crucial property that they are eigenstates of the annihilation and creation operators, i.e., 
    \begin{align}\label{eq-coherent-state-eigenvalue}
        a_\ell\eta_z \, = \, z_\ell\eta_z \, .
    \end{align}
    If $\psi\in \cQ_\enum(B)$ is normalized, observe that $\psi_z = \scp{\eta_z}{\psi}_{\cF (M)}\in \big(\bigwedge_{j=1}^nL^2(B)\big)\otimes\cF(M^{\perp})$. For notational simplicity, we will now omit the Hilbert space in the inner products. Using that 
    \begin{align}
        \cF(M)\cong \bigoplus_{(\ell,n)\in \Lambda_P\times\N_0} \Span\{(\ad_\ell)^n \Omega\} \, ,
    \end{align}
    a calculation in the sense of weak integrals on the Fock space $\cF (M)$ with measure $\dx{z}= \prod_{\ell\in\Lambda_P}\frac{1}{\pi}\int \dxx{x_\ell}\dx{y_\ell}$ gives
	\begin{align}
		&\int \scp{.}{\eta_z}\eta_z \dx{z} \, = \, 1, &&\int  z_\ell\scp{.}{\eta_z}\eta_z \dx{z} \, = \, a_\ell,\nonumber\\
		&  \int  \overline{z}_\ell\scp{.}{\eta_z}\eta_z \dx{z}\, = \, \ad_\ell, &&\int (|z_\ell|^2-1)\scp{.}{\eta_z}\eta_z \dx{z} \, = \, \ad_\ell a_\ell,\label{identities}
	\end{align}
	where the second and third equalities follow from the first identity, using \eqref{eq-coherent-state-eigenvalue}, and the last then follows from the fact $[a_\ell,\ad_\ell]=1$ for all $\ell\in\Lambda_P$. Let the block modes be replaced by the identities \eqref{identities}, then
	\begin{align}
		\scp{\psi}{H_{\mathrm{block}}^{(\enum,\alpha,\beta)}\psi} = \beta \int \scp{\psi_z}{h_z\otimes 1\psi_z}\dx{z} ,\label{int:hz}
	\end{align}
	whereas $h_z$  is a Schr\"odinger operator on $L^2(\R^{3\enum})$
	\begin{align*}
		h_z &=\sum_{j=1}^\enum \Big(D_{A_\alpha,x_j}^2+\beta^{-1}V_\alpha(x_j)+\frac{\sqrt{\alpha}}{\sqrt 2\pi\beta}\sum_{\ell\in\Lambda_P}M_\ell\left(z_\ell e^{ik_\ell x_j}+\overline{z}_\ell e^{-ik_\ell x_j} \right)\\
        &\qquad +\sum_{k=j+1}^N\frac{\nu\alpha\beta^{-1}}{|x_j-x_k|}\Big) + \sum_{\ell\in\Lambda_P}(|z_\ell|^2-1).
	\end{align*}
	Denoting $\rho_z(x) :=\frac{\enum}{\|\psi_z\|^2}\int_{\R^{3(\enum-1)}}|\psi_z(x,X)|^2\dx{X}$, we find that
	\begin{equation}
	   \sum_{j=1}^\enum \scp{\psi_z}{e^{-ikx_j}\psi_z} = n \scp{\psi_z}{e^{-ikx_1}\psi_z} \, = \, \|\psi_z\|^2 \hat{\rho}_z (k) \, .    
	\end{equation}
    Recalling \eqref{def-block-modes}, we find that
    \begin{align}\label{eq-rho-FT-block-est}
        -\inf_{k_\ell\in Q(\ell)}|\hat{\rho}_z (k_\ell)|^2M_\ell^2 \, \geq \, - \int_{Q(\ell)}\frac{|\hat{\rho}_z(k)|^2}{|k|^2}\dx{k} \, .
    \end{align}
	By completion of squares, we thus have that 
    \begin{align}
        \MoveEqLeft
        \sup_{k_\ell\in Q(\ell)}\scp{\psi_z}{\Big(|z_\ell|^2+\frac{\sqrt{\alpha} M_\ell}{\sqrt{2}\pi\beta}\sum_{j=1}^\enum \left(z_\ell e^{ik_\ell x_j}+\overline{z}_\ell e^{-ik_\ell x_j} \right) \Big)\psi_z} \\ 
        &= \, \|\psi_z\|^2\Big(\Big|z_\ell+\frac{\sqrt{\alpha}M_\ell \hat{\rho}_z(k_\ell)}{\sqrt{2}\pi\beta}\Big|^2-\frac{\alpha}{2\pi^2\beta^2} \inf_{k_\ell\in Q(\ell)}|\hat{\rho}_z (k_\ell)|^2M_\ell^2\Big)\\
        &\geq \, -\frac{\alpha\|\psi_z\|^2}{2\pi^2\beta^2}\int_{Q(\ell)}\frac{|\hat{\rho}_z(k)|^2}{|k|^2}\dx{k} \, .
    \end{align}
    As a consequence, employing the fact that $\sum_{\ell\in\Lambda_P}\int_{Q(\ell)}=\int_{B_\Lambda(0)}$, we find that
	\begin{align}
		\MoveEqLeft \sup_{\{k_\ell\}}\int \langle\psi_z,h_z\psi_z \rangle\dx{z} \\
		&\geq \, \int \langle \psi_z,\sum_{j=1}^\enum\Big(D_{A_\alpha,x_j}^2+ \beta^{-1}V_\alpha(x_j)+\sum_{k=j+1}^N\frac{\nu\alpha\beta^{-1}}{|x_j-x_k|}\Big)\psi_z \rangle \dx{z} \\
        &\qquad -\frac{\alpha}{2\pi^2\beta^2}\int \|\psi_z\|^2\int_{\R^3}\frac{|\hat{\rho}_z (k)|^2}{|k|^2}\dx{k}\dx{z}-|\Lambda_P|\\
		&= \, \int  \langle \psi_z,\sum_{j=1}^\enum\Big(D_{A_\alpha,x_j}^2+ \beta^{-1}V_\alpha(x_j)+\sum_{k=j+1}^N\frac{\nu\alpha\beta^{-1}}{|x_j-x_k|}\Big)\psi_z \rangle\dx{z}\\ \label{eq-ineqintegrant}
        & \qquad -\frac\alpha{\beta^2}\int \|\psi_z\|^2\int_{B}\int_{B}\frac{\rho_z(x)\rho_z(y)}{|x-y|}\dx{x}\dx{y}\dx{z}-|\Lambda_P| \, ,
	\end{align}
    where we used the fact that the Fourier transform of $|k|^{-2}$ satisfies $(2\pi)^{-3}\int\frac{e^{ikx}}{|k|^2}\dx{k}=\frac{2\pi^2}{|x|}$ in the distributional sense. Using \eqref{def-EPT-concave} and \eqref{eq-ineqintegrant} we find
	\begin{align}
       	\MoveEqLeft \sup_{\{k_\ell\}}\int \langle\psi_z,h_z\psi_z \rangle\dx{z} \geq  \EPT^{(\enum,\alpha,\beta^{-1})}(B) \int \|\psi_z\|^2 \dx{z}=\EPT^{(\enum,\alpha,\beta^{-1})}(B) \, ,
	\end{align}  which together with \eqref{int:hz} implies
	\eqref{eq:pekar}.
    
\end{proof}

\end{document}